\documentclass[aps,prl,twocolumn,floatfix]{revtex4}

\usepackage{hyperref}
\usepackage{graphicx}
\usepackage[utf8]{inputenc}
\usepackage{amsmath}
\usepackage{comment}
\usepackage{bm}
\usepackage{dsfont}
\usepackage{amsfonts}
\usepackage{xcolor}

\usepackage{braket}
\usepackage{siunitx}
\usepackage{tabularx}
\usepackage{float}
\usepackage{braket}
\usepackage{placeins}

\begin{document}

\title{Remote spin control  in Haldane spin chains}

\author{Y. del Castillo$^{1,2}$, A. Ferr\'on $^3$, J. Fern\'{a}ndez-Rossier$^1$\footnote{On permanent leave from Departamento de F\'{i}sica Aplicada, Universidad de Alicante, 03690 San Vicente del Raspeig, Spain}$^,$\footnote{joaquin.fernandez-rossier@inl.int} }

\affiliation{$^1$International Iberian Nanotechnology Laboratory (INL), Av. Mestre Jos\'{e} Veiga, 4715-330 Braga, Portugal }
\affiliation{$^2$Centro de F\'{i}sica das Universidades do Minho e do Porto, Universidade do Minho, Campus de Gualtar, 4710-057 Braga, Portugal }
\affiliation{$^3$Instituto de Modelado e Innovación Tecnológica (CONICET-UNNE) and Facultad de Ciencias Exactas, Naturales y Agrimensura, Universidad Nacional del Nordeste, Avenida Libertad 5400, W3404AAS Corrientes, Argentina. }

\date{\today}

%%%%%%%%%%%%%%%%%%%%%%%%%%%%%%%%%%%%%%%%%%%%%%%%%%%%%%%%%%

\begin{abstract}
We consider the remote manipulation of the quantum state of the edge fractional spins of Haldane spin chains using a weak local perturbation on the other edge. We derive an effective four-level model that correctly captures the response of the  local magnetization to local perturbations and we use it to show that applying a small local field on one edge of the chain induces a strong variation of the magnetization on the opposite edge. Using a Landau-Zener protocol, we show how local control of the field on one edge of the chain, implemented for instance with a spin-polarized scanning tunnel microscope tip, can adiabatically switch the magnetization direction on the other side of the chain.
\end{abstract}

\maketitle

Haldane spin chains\cite{haldane83a,haldane83b} with periodic boundary conditions have an $S=0$ ground state and a gapped spin excitation spectrum, making them different from 
$S=1/2$ Heisenberg chains, that feature a gapless excitation spectrum\cite{Clozieaux62}. This difference is even more striking in the case of open-end chains, as the lowest energy levels of the  Haldane chains are a quasi-degenerate singlet-triplet quartet, associated with the emergence of two effective $S=1/2$ degrees of freedom localized in a few sites at the edges of the chains\cite{kennedy90}.  The  splitting of the singlet-triplet quartet $j$ decays exponentially with the chain size,
$j\simeq J e^{-N/\xi}$
so that, effectively, in the thermodynamic limit $N\gg\xi$, the two effective edge spins become independent.  However, there is a mesoscopic range, where $N/\xi$ is not large and temperature is small where the effective coupling $j>>k_BT$, so that the system occupies, with large probability, 
a non-degenerate singlet ground state where the two emergent edge $S=1/2$ spins, physically separated,  are entangled.

Here, we address the question of whether it would be possible to leverage this non-local entanglement in Haldane spin chains to achieve a non-local manipulation of the effective spin in one edge of the Haldane chain by acting locally on the opposite edge. We refer to this as remote manipulation.
The idea of using spin chains as channels to transfer information, either quantum or classical, has been thoroughly explored theoretically\cite{Bose2003,Christandl04,Burgarth2005,bose2007,campos07,romero07,zwick11,nikolopoulos2014,banchi17,bazhanov18,van21,Coden2024}. The experimental demonstration of these ideas has only become possible with the advent of physical platforms that allow to fabricate and probe individual spin chains, using as building blocks magnetic adatoms\cite{hirjibehedin06,khajetoorians11,choi19}, nanographenes\cite{Mishra2021,zhao24,zhao25}, quantum dots\cite{vandiepen21}, donors in silicon\cite{donnelly24},  cold-atoms\cite{Murmann2015}, trapped ions\cite{senko15} and Rydberg atoms\cite{barredo15}. Transfer of classical information across an adatom spin chain was demonstrated experimentally\cite{khajetoorians11}, and transmission, control, and manipulation of quantum states have been achieved using quantum dots\cite{Kandel2021}, cold atoms\cite{ Murmann2015}, and magnetic atoms on surfaces\cite{Wang2023}.

The specific case where the end-states of Haldane spin chains are exploited to enhance their performance as information channels seems to remain unexplored. Some works have considered the role of quantum phase transitions to enhance the channel capacity of $S=1$ Haldane chains\cite{romero07,banchi17}. There are at least two different realizations of the Haldane phase for which the discussion of this work is relevant. First, the original formulation of Haldane\cite{haldane83a,haldane83b}, with $S=1$ antiferromagnetically coupled spins, that may or may not have a non-linear exchange $\beta$. Second, the alternate exchange $S=1/2$  Heisenberg model, with two coupling constants $J_1$, $J_2$, where at least one of them has to be antiferromagnetic\cite{hida92}. There are several physical platforms where {\em individual} Haldane spin chains, made either with $S=1$ and constant exchange or with $S=1/2$ and alternate exchange, can be implemented or are expected to be realizable: magnetic nanographenes\cite{Mishra2021,zhao24}, magnetic adatoms\cite{wang24}, cold atoms\cite{sompet22}, and phosphorous dopants in silicon\cite{munia24}.   Very similar physics can also be 
implemented with cold hard-core bosons\cite{de19}. The implementation of Haldane spin chains has also been proposed using Rydberg atoms\cite{mogerle24} and quantum dots\cite{jaworowski17,baran24}.
%\bluemark{In addition, there are several classes of compounds with weakly coupled one-dimensional spin chains that feature Haldane physics\cite{buyers86,tun90,brunel92,bertaina14}}

%In all these systems, the distance between the edge spins is governed by the physical separation between adjacent spins, $d$, and the magnitude of the spin exchange.  
%In the case of nanographene triangulenes, intermolecular exchange is very large \cite{Mishra2021}  ($J\simeq 19$ meV) and $d\simeq 1nm$, given by the size of triangulenes. 
%This leads to a Haldane gap of approximately 8 meV, and a sizable inter-edge exchange $j\simeq 0.8\mu$eV for $N=40$ and therefore, large inter-edge distance, in the range of $N\cdot d\simeq 40 nm$. 

\begin{figure}[t]
    \centering
    \includegraphics[width=1\linewidth]{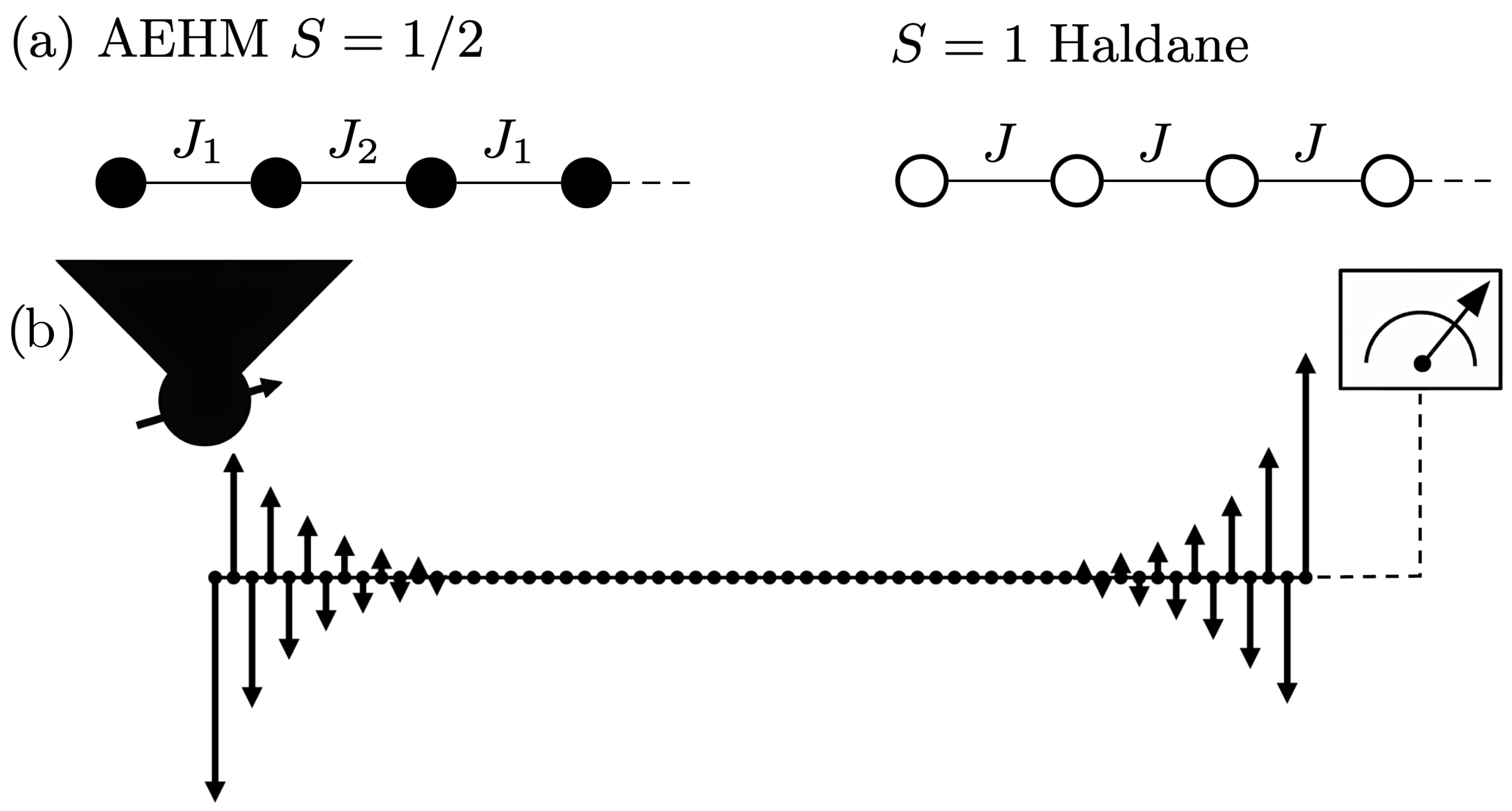}
    \caption{ (a) Alternating exchange Heisenberg model (AEHM) chain with S=1/2 (left), S=1 Haldane spin chain (right). (b) Schematic representation of a spin chain where a local magnetic field is applied to the first spin, and its resulting magnetization is measured on the last spin. }
    \label{fig:scheme}
\end{figure}
%S=1 Haldane chain with N=60 and $\beta=0.09$ computed with density matrix renormalization group (DMRG) code dmrgpy\cite{lado2023dmrgpy}.
Having in mind the case of nanographenes\cite{Mishra2021,zhao24}, we thus consider two spin-rotational invariant Hamiltonians, build
with the Heisenberg coupling $h_{i,j}=\vec{S}_i\cdot\vec{S}_{j}$. The first one is the $S=1$ Haldane spin chain model\cite{Affleck87} (See Fig. \ref{fig:scheme}(a) right panel):
\begin{equation}
    \begin{split}
    {\cal H}^{(S=1)}_{H}= \sum_{n=1,N-1} J & \bigg[h_{n,n+1}+\beta h_{n,n+1}^2
    %\vec{S}_n\cdot  \vec{S}_{n+1}+\beta\left(\vec{S}_n\cdot %\vec{S}_{n+1} \right)^2
    \bigg ] 
    %\\
    % &+\sum_{n=1,N} g\mu_B \vec{S}_n\cdot\vec{B}
    \end{split}
\label{eq:ham1}
\end{equation}
where we consider $J>0$, to ensure antiferromagnetic interactions, and $0\leq \beta \leq \frac{1}{3}$, to ensure a gap $\Delta_{\rm H}$ in the excitation spectrum in the case with periodic boundary conditions (PBC). 
The second case is the alternate exchange Heisenberg model\cite{diederix79,bonner83,hida92} (AEHM) for $S=1/2$ (See Fig. \ref{fig:scheme}(a) left panel):
\begin{equation}
    {\cal H}^{(S=1/2)}_H=
    %\sum_{v=0,1} J_{1+v}
     %   \sum_{n=0,N_{max}}  \vec{S}_{2n+v}\cdot\vec{S}_{2n+1+v}
     \sum_{n=0,\frac{N-2}{2}}  \left(J_1 h_{2n,2n+1} + J_2  h_{2n+1,2n+2}\right)
 %   +
 %   J_2 \sum_{n=0,\frac{(N-2)}{2}}  \vec{S}_{2n+1}\cdot\vec{S}_{2n+2} 
\label{eq:ham2}
\end{equation}
In both cases, we take $N$ as an even number, ensuring that the ground state has $S=0$, 
and $N_{max}=\frac{(N-2)}{2}$. 
The AEHM also features a gap  $\Delta_{\rm H}$ for PBC and $J_1\neq J_2$ when at least one of the two exchange parameters is positive. 
We note that the gapped phases of AEHM are adiabatically connected to the Haldane model\cite{hida92}, by making either $J_1$ or $J_2$ negative and very large. Therefore, we refer to the gapped phases of both the $S=1$ model and the AEHM with antiferromagnetic exchange and $J_1\neq J_2$ as the Haldane phase.

In the Haldane phase, chains with open boundary conditions (OBC) feature four low energy levels, formed by a singlet $S$ and a triplet $(T_0, T_+,T_-)$, reflecting the emergence of $S=1/2$ edge spins. The  singlet-triplet splitting can be written as $j\simeq J e^{-N/\xi_1}$ in the $S=1$ chains, and as $j\simeq J_1 e^{-(N-2)/\xi_{1/2}}$ in the $S=1/2$.  Hence, in the thermodynamic limit and OBC, both models have a fourfold degenerate ground state. 

Here we are interested in {\em finite-size} chains  where  $k_BT<< j<<\Delta_{\rm H}$, so 
that only the ground state manifold states are occupied, so that they are the ones that
control the response of the system. We  can thus derive  an analytical theory, valid for  large spin chains, to model  the effect of the application of 
  a local magnetic field  $b$ coupled to the outermost spin on one edge of the chain.  
  We consider the following Hamiltonian: 
  \begin{equation}
    {\cal H}={\cal H}^{(S)}_H + {\cal V}= {\cal H}_H^{(S)}+ g\mu_B b \hat{S}^z_1 
\end{equation}
The local field $b$ could be induced, for instance, through local exchange with an STM tip\cite{yang19}, as well as the exchange field with a nearby magnetic atom or molecule\cite{phark23}.  The discussion that follows would be the same if the local field acts on a few spins on one edge, instead of only one.  We assume that  the effect of the local field $b$ 
 is {\em small} compared to the size of the Haldane gap, $\Delta_{\rm H}$, an assumption easy to meet in the case of nanographene spin chains.     
Therefore, we treat the effect of the local field in the first spin using first-order degenerate perturbation theory. %which makes the analysis very similar 
%to our recent work \cite{}.
To do so, we label the states of the singlet-triplet ground state manifold with $G=\left(S,T_0, T_+,T_-\right)$.
To leading order, the   effective Hamiltonian in the $G$ subspace reads
\begin{equation}
    {\cal H}_{eff}(b) =  \langle G|{\cal V}|G'\rangle 
%    + \sum_X \frac{\langle G|{\cal V}|X\rangle\langle X|{\cal V}|G'\rangle}{E_G-E_X}
\end{equation}
%We have tested the validity of this approximation by comparing it with exact numerical diagonalization. (Decimos lo mismo mas abajo)

In order to write the effective Hamiltonian matrix, we note that 
the expectation value of the local spin $S^z_1$, computed in the ground state manifold, vanishes for $S,T_0$, as these are states with $S^z=0$.
Therefore, the only non-null matrix elements for the local spin operators are:
\begin{equation}
    {\cal S}^z_i \equiv \langle S|\hat{S}^z_i |T_0\rangle, 
    \;\;    {\cal T}^{(\pm)}_i \equiv \langle T_{\pm}|\hat{S}^z_i |T_{\pm}\rangle
    \label{eq:LSO}
\end{equation}
%and
%\begin{equation}
%    {\cal T}^{(\pm)}_i \equiv \langle T_{\pm}|\hat{S}^z_i |T_{\pm}\rangle
%\end{equation}

\begin{figure}[h!]
    \centering    \includegraphics[width=1\linewidth]{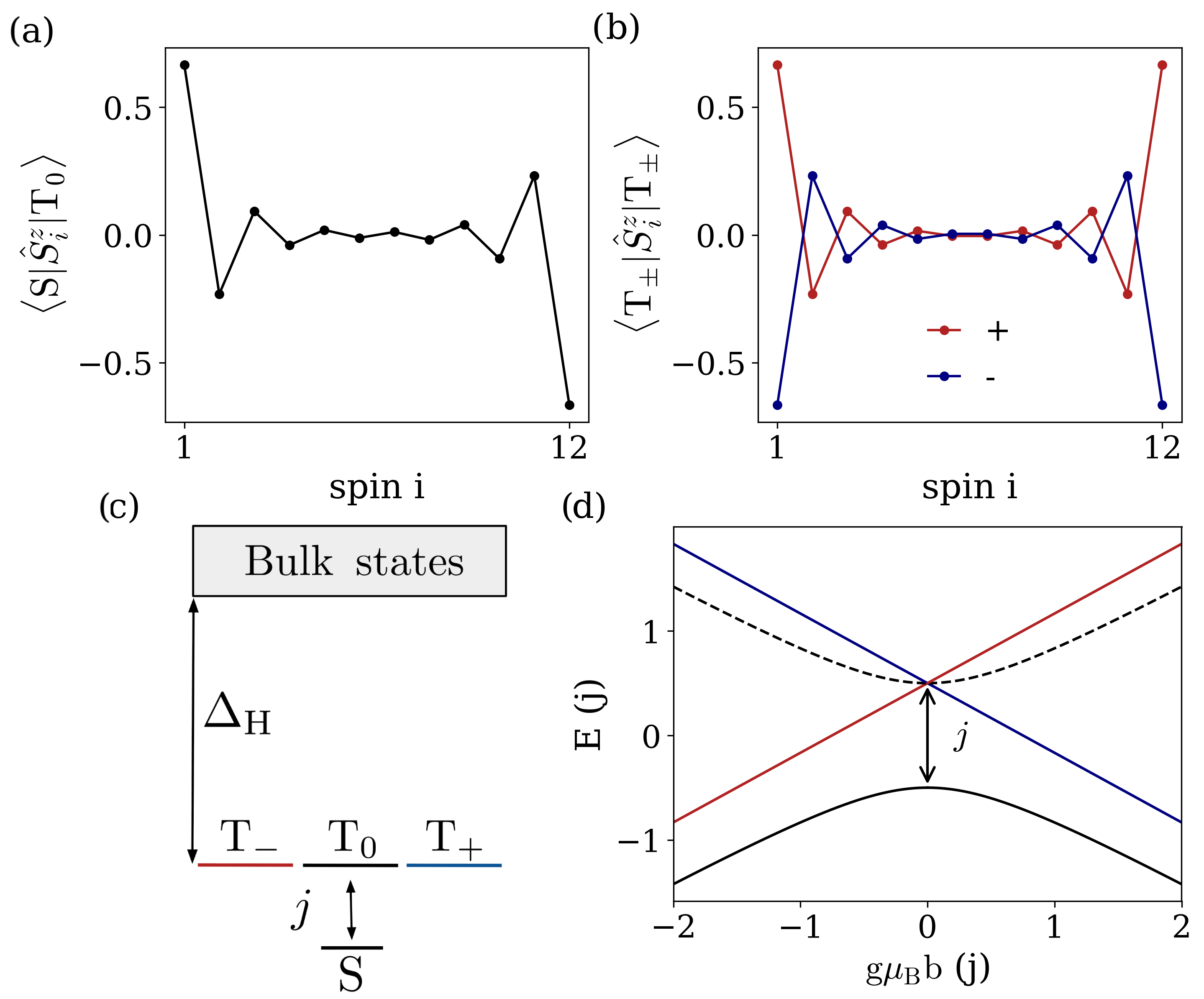}
    \caption{Haldane S=1 non-vanishing matrix elements of the local spin operators, (a) ${\cal S}_i^z$ and (b) ${\cal T}_i^{(\pm)}$, for each spin with no local field applied. (c) Energy level scheme with the Haldane gap, $\Delta_{\rm H}$, separating bulk states and the low-energy manifold, and the singlet-triplet gap, $j$. (d) Energy diagram of the ground state manifold in the S=1 Hamiltonian, Eq. (\ref{eq:ham1}), as a function of the local magnetic field applied to the first spin, expressed in units of the singlet-triplet splitting $j$. Simulation parameters: N=12, $\beta$=0.3, resulting in $j=2.63\times 10^{-4}$ J). }
    \label{fig:EffectiveModel}
\end{figure}

The matrix elements ${\cal S}^z_i$ and ${\cal T}^{(\pm)}_i$ play a central role in the rest of the paper. In Figs.~\ref{fig:EffectiveModel}(a) and \ref{fig:EffectiveModel}(b), we plot them both respectively for the $S=1$ chain. They peak at the edges, and ${\cal S}^z_i$ has antiferromagnetic inter-edge correlation. 
We note that these matrix elements are computed with the unperturbed Hamiltonian and are therefore independent of $b$. The dependence on $b$ enters via the prefactors in the ${\cal V}$ operator:
\begin{equation}
    \epsilon_0(b)\equiv  g\mu_B b  {\cal S}^z_1, \;\;
    \epsilon_{\pm}(b)\equiv g\mu_B b  {\cal T}^{(\pm)}_1 
\end{equation}
%and
%\begin{equation}
%    \epsilon_{\pm}(b)\equiv g\mu_B b  {\cal T}^{(\pm)}_1 
%\end{equation}
We find that $|{\cal T}^{(\pm)}_1|= | {\cal S}^z_1|$. We thus write:
\begin{eqnarray}
    {\cal H}_{eff}(b) = \left(
    \begin{array}{cccc} 
  -\frac{j}{2} & \epsilon_0(b)   & 0 & 0 \\
  \epsilon_0(b) &   \frac{j}{2}& 0   & 0 \\
0 & 0 & \frac{j}{2}+\epsilon_+(b) & 0 \\
0 & 0&0 & \frac{j}{2}+\epsilon_-(b)
    \end{array}
    \right)
    \label{eq:4x4}
\end{eqnarray}
%\bluemark{
This Hamiltonian is analogous to the one describing singlet-triplet qubits\cite{petta05} and edge spin states in a rectangular zigzag ribbon\cite{ortiz18}.  
%}.
Thus, we see that the two states of the $S^z=\pm 1$ sector, with energies $\frac{j}{2}+\epsilon_\pm(b) $, are decoupled from those of the $S^z=0$ sector, that are described by an effective two-level model  in the $S,T_0$ subspace:
\begin{equation}
    h(b) = -\frac{j}{2}\tau_z + \epsilon_0(b)\tau_x = \vec{b}(b) \cdot\vec{\tau}
    \label{eq:TLS}
\end{equation}
where $\tau_x,\tau_z$ are Pauli matrices
%$\tau_z= \left(\begin{array}{cc} 1 & 0 \\0 & -1 \end{array}\right)$
%and $\tau_x= \left(\begin{array}{cc} 0 & 1 \\ 1 & 0 \end{array}\right)$, 
and $\vec{b}=\left(\epsilon_0(b),0,-\frac{j}{2}\right) $.
The eigenvalues of this two-level system are given by $h(b) |\psi_\pm(b) \rangle = \pm E(b) |\psi_\pm(b) \rangle$, with
\begin{equation}
    E(b)= \frac{1}{2}\sqrt{j^2 + 4\epsilon_0(b)^2}
\end{equation}
In Fig. \ref{fig:EffectiveModel}(d), we plot the energies of the 4 low energy states of the Hamiltonian \ref{eq:4x4}, for a spin chain with N=12 and $\beta=0.32$. In the Supplemental Material\nocite{ITensor, lado2023dmrgpy, del24}\cite{supp}, we show that exact numerical diagonalization of this effective model yields identical results for both the $S=1$ and the AEHM $S=1/2$ models using Eq. (\ref{eq:ham1}) and \ref{eq:ham2}, respectively. It is apparent that the state $\psi_-$ of the $S^z=0$ manifold is always the ground state.
The physical properties of the $S^z=0$ manifold are governed by the eigenstates:
%\begin{eqnarray}
 %   |\psi_-\rangle=-\sin\frac{\theta(b)}{2}|T_0\rangle + \cos\frac{\theta(b)}{2} |S\rangle
%\end{eqnarray}
%\begin{eqnarray}
 %   |\psi_+\rangle=\cos\frac{\theta(b)}{2}|T_0\rangle + \sin\frac{\theta(b)}{2} |S\rangle
%\end{eqnarray}
%
\begin{eqnarray}
\left(\begin{array}{c}  \ket{\psi_-} \\ \ket{\psi_+}\end{array}\right) =  
\left(\begin{array}{cc} 
\cos\frac{\theta(b)}{2} & -\sin\frac{\theta(b)}{2} \\
\sin\frac{\theta(b)}{2} &  \cos\frac{\theta(b)}{2} 
\end{array}\right)
\left(\begin{array}{c}  \ket{S} \\ \ket{T_0}
\end{array}\right)  
\label{eq:states}
\end{eqnarray}
\noindent where $\cos \theta(b)= \frac{-j}{2E(b)}$, $\sin\theta(b) = \frac{\epsilon_0(b)}{E(b)}$.  From  Eq. (\ref{eq:states}), it is straightforward to obtain  the 
local magnetization in response to a field applied on site $1$ of the chain: 
\begin{equation}
    \langle \psi_{\pm}|\hat{S}^z_i |\psi_{\pm}\rangle= %\pm \sin \theta(b) {\cal S}^z_i=
    \frac{\pm2\epsilon_0(b){\cal S}^z_i}{\sqrt{4\epsilon_0(b)^2 + j^2}} 
    =\frac{\pm2g\mu_B {\cal S}^z_1{\cal S}^z_i b}{\sqrt{4\epsilon_0(b)^2 + j^2}}
    \label{eq:mag}
\end{equation}
\begin{figure}[h]
    \centering
   \includegraphics[width=1\linewidth]{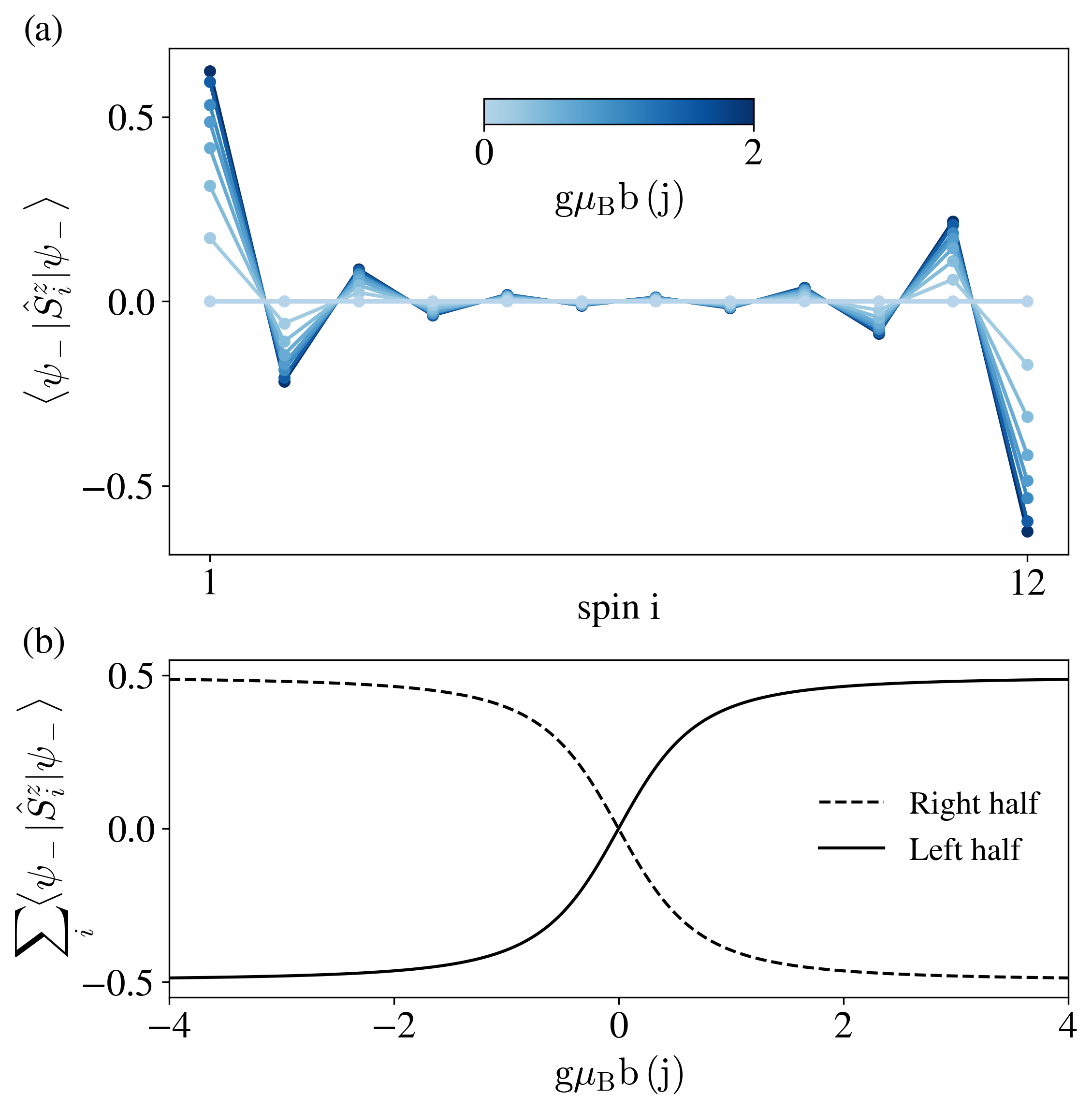}
    \caption{(a) Ground-state expectation values $\langle S^z_i \rangle$ obtained from the full Hamiltonian, Eq. (\ref{eq:ham1}), for different local fields $g\mu_B b$ applied at the first spin, $i=1$. (b) Total magnetization of the left and right halves of the chain, computed with the effective model \ref{eq:mag}, as a function of the local field applied on the first spin. Parameters: $N=12$ and $\beta=0.32$.}
    \label{fig:magnetization-chain}
\end{figure}
This is the central result of this paper. It shows how the application of a local magnetic field in just one edge spin of the chains described by the Haldane spin chains, described by Hamiltonians \ref{eq:ham1} and \ref{eq:ham2}, affects the magnetic state of the entire chain.  Given the very large antiferromagnetic inter-edge spin correlation, it affects the opposite edge with equal intensity, and opposite sign, on account of the profile of the matrix element ${\cal S}^z_i$. Equation (\ref{eq:mag}) shows that the key figure of merit that controls the average local magnetization of the Haldane spins 
is the ratio $\frac{\epsilon_0(b)}{j}$. 
For $b=0$ the magnetization of the entire chain vanishes, as expected for the $S^z=0$ states of the ground state manifold. In the opposite limit, $|\epsilon_0|>>j$, 
 we have
\begin{equation}
    \langle \psi_{\pm}|\hat{S}^z_i |\psi_{\pm}\rangle\simeq \pm {\cal S}^z_i
\end{equation}
so that both edges have a local magnetization of order 1$\mu_B$ (see Figs.~\ref{fig:magnetization-chain}(a) and \ref{fig:magnetization-chain}(b)), with opposite magnetization at both edges. Thus, in this limit, a local perturbation in one edge is able to saturate the magnetization on the opposite edge.  

%\bluemark{Hamiltonians \ref{eq:4x4} and \ref{eq:TLS}}, together with 
The extremely non-local character of the spin response of the Haldane chains opens the door for  different applications. First, as a consequence of \ref{eq:TLS},  application of an AC local field $b(t)=b_0 \cos\omega t$ will result in resonant excitation of transitions between the singlet and triplet when $\hbar\omega=j$.  Thus, a local AC field  could provides a driving mechanism for  electron spin resonance with STM \cite{baumann15}  in individual Haldane spin chains\cite{maiellaro24}.   Second,  for non-local sensing: a field acting on one edge of the chain has an impact on the other edge, where a local probe could be placed (see Fig.~\ref{fig:scheme}(b)). Third,  non-local adiabatic manipulation of the edge magnetization. Equation \ref{eq:4x4} shows that the $S^z=0$ manifold of the Haldane chains with a local magnetic field is described with a two-level Hamiltonian with an avoided level crossing. Therefore, adiabatic control of the ground-state wave function would allow dynamic control of the magnetization at one edge by applying a local field at the other.

\begin{figure}[h!]
    \centering
\includegraphics[width=1\linewidth]{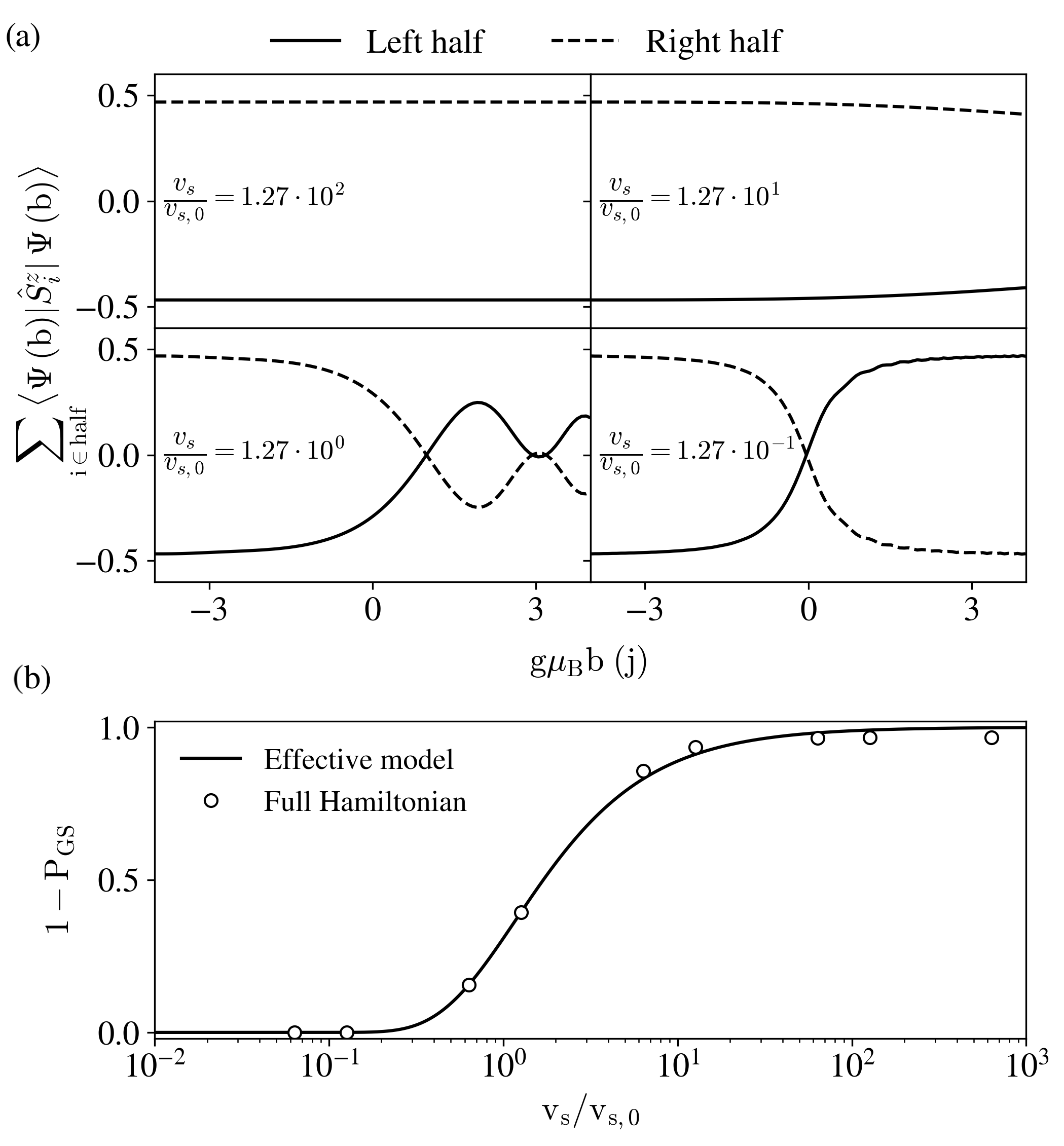}
    \caption{
(a) Ground–state magnetization of the left (solid) and right (dashed) halves as a function of $g\mu_B b/j$ for four sweep rates using Eq. (\ref{eq:ham1}). 
(b) Landau–Zener non-adiabatic transition probability, $1-P_{\mathrm{GS}}$, as a function of the sweeping rate $v_s/v_{s,0}$, $v_{s,0}=j^2/(\hbar g\mu_B)$. The black curve corresponds to the effective model and the white dots are numerical results using the complete Hamiltonian, Eq. (\ref{eq:ham1}). Parameters: $N=8$, $\beta=0.32$.
}
    \label{fig:sweep-velocity}
\end{figure}

To illustrate this, in Fig. \ref{fig:sweep-velocity}(a), we simulate a Landau–Zener sweep in an $S=1$ Haldane spin chain of $N=8$ sites with $\beta=0.32$. A local magnetic field $b$ applied to one edge of the chain is ramped linearly as $b_1(t) = b_0 + v_s t$, where the sweep rate $v_s = \frac{db}{dt}$ sets the driving speed and, at the avoided crossing, induces Landau–Zener (LZ) transitions.
%To illustrate this, we simulate a Landau–Zener sweep in an $S=1$ Haldane spin chain of $N=8$ sites with $\beta=0.32$. A local magnetic field $b$ acting on one edge of the chain is ramped following a linear law $b_1(t)=b_0+v_s t$ where the sweep rate or the varying velocity of the magnetic field $v_s=\frac{db}{dt}$ is the constant that controls the pace of the driving and induces, at the avoided crossing, Landau-Zener (LZ) transitions.
The initial point must be positioned far from the avoided crossing. %\sout{
%To achieve this, we can either apply a local magnetic field $b_0$ to set the desired initial position or turn off this magnetic field and begin at an initial time different from zero}
%\redmark{(estos detalles se pueden obviar pero debemos decidir que hacemos. En mis calculos yo pongo un $b_0$ negativo para arrancar a la izq del cruce.)}.
%Both approaches yield the same effect if $t_0=b_0/v_s$.

We first model the Landau–Zener sweeps using the full numerical solution of the Hamiltonian in Eq. (\ref{eq:ham1}). At $t_i = 0$, we take the system to be in its ground state for a negative local field $g\mu_B b_0 = -4j$, and we compute the dynamical evolution of the quantum state as $b$ is ramped so that, at $t = t_f$, $g\mu_B b = +4j$.
The time it takes for the LZ- sweep is given by $\Delta t_{LZ}=\frac{\Delta b}{v_s}$.
%\bluemark{We repeat this procedure for various values of the sweep rate $v_s$ and we compute the state $|\Psi(t)\rangle=|\Psi(b)\rangle$ at the end of the sweep and evaluate the average magnetization of each half of the chain, $\sum_{i\in {\rm half}} \bra{\Psi(b)} \hat S_i^z\ket{\Psi(b)}$.}
At each time step, we compute the state $|\Psi(t)\rangle=|\Psi(b)\rangle$ of the sweep and evaluate the average magnetization of each half of the chain, $\sum_{i\in {\rm half}} \bra{\Psi(b)} \hat S_i^z\ket{\Psi(b)}$. We repeat this procedure for various values of the sweep rate $v_s$.
We focus on the probability of exciting the system out of the adiabatic ground state during the sweep, as a function of the sweep velocity $v_s$. If the system remains in the ground state, the magnetization is reversed. On the contrary, if the system undergoes a transition from the ground to the excited state, the magnetization remains constant throughout the sweep.

Using the only intrinsic energy scale of the low-energy spectrum, 
$j$, we define the {\em natural} scale for the magnetic sweep rate,  $v_{s,0}\equiv\frac{j^2}{\hbar g\mu_b }$.  
We find that, for large sweep velocities, $v_s>>v_{s,0}$, the magnetization remains constant during the sweep of the local field. This clearly implies that the quantum state of the system, $|\Psi(t_f)\rangle$, is no longer in the adiabatic ground state for $g\mu_B b_1>>j$. In the opposite limit, when $v_s \ll v_{s,0}$, the magnetization is reversed during the Landau-Zener sweep, which reflects an adiabatic evolution.  We also show two intermediate situations.

In Fig. \ref{fig:sweep-velocity}(b) we compare our numerical results, obtained for a chain with $3^8=6561$ states, to the Landau-Zener formula\cite{Landau1932,Zener1932,ivakhnenko2023} for the probability of exciting the system out of the adiabatic ground state during the local field sweep, valid for our two-level model of Eq. (\ref{eq:TLS}):
\begin{equation}
%P_{LZ}=\exp{\left[-\frac{\pi j^2}{4\hbar g\mu_B {\cal S}^z_1 v_s}\right]}
P_{LZ}=\exp{\left[-\frac{\pi v_{s,0}}{4 {\cal S}^z_1 v_s}\right]}\,.
\label{LZ}
\end{equation}
We find a very good agreement between the full numerical calculation and the LZ model. This extends the validity of the effective model to the time-dependent case. More importantly, it showcases the potential of Haldane spin chains as platforms to carry out non-local adiabatic manipulation of distant quantum states.

We now discuss the experimental conditions required so that the approximations used for our model can be applied. These can be summarized in the following inequalities:
\begin{equation}
    k_{\mathrm{B}}T <% g\mu_B B_{\rm ext}< 
    j < g\mu_{\mathrm{B}}b << \Delta_{\mathrm{H}} .
\end{equation}
%The applicability of the effective model requires a clear hierarchy of energy scales. 
the correlation lengths  $\xi_{1}$  and  $\xi_{1/2}$ that, in turn, are controlled by the dimensionless parameters   $\beta$  and $\delta = \frac{J_1-J_2}{J_1+J_2}$, respectively.
In the case of $S=1$ nanographenes\cite{Mishra2021},  the relevant values are $\beta=0.09$, $J=19$meV that, according to our DMRG calculations, give $\xi_1\simeq4.13$  and $\Delta_H=7.8$ meV, for the $S=1$ chains. In turn, for the $S=1/2$ AHEM, the relevant values are 
$J_1=21$ meV, $J_2=37$ meV, 
$\delta \simeq -0.28$; that yield $\xi=2.41$  $\Delta_H=22$ meV
Hence, the condition $j<<\Delta_{\mathrm{H}}$  is easily met by increasing chain length.  
For instance, for $S=1$ nanographene chains\cite{Mishra2021}, the singlet–triplet gap for a chain of $N = 22$ spins is of the order of $j\simeq 94 \mu$eV. 

%We require that the Haldane gap is much larger than all the other energy scales in the problem to ensure that the system dynamics remain restricted to the low-energy manifold, without  populating higher-energy states.  In any Haldane system, $j$ decays exponentially with length, hence having $\Delta_{\mathrm{H}}>>j$ can be easily achieved  by using long chains.

The condition $j < g\mu_{\mathrm{B}}b$ is required to fully polarize the edge spins (see Fig. \ref{fig:magnetization-chain} and Eq. (\ref{eq:mag})), 
the Zeeman energy of the local field must be at least two times $j$,  $g\mu_{\mathrm{B}}b > 2j$, a total $\Delta g\mu_{\mathrm{B}}b = 4j$ for the complete sweep. For the above value of $j\sim 94 \mu$eV, this translates into a $\Delta b \sim  3.2$ T, that can easily be achieved, for instance, with a scanning tunnel microscope (STM) tip\cite{yang19}.
The %scales for both the external magnetic field and the
thermal energy must be at least one order of magnitude smaller (see Supplemental Material\cite{supp} for a numerical analysis of the ground state thermal occupation), corresponding to %$B_{ext}\sim 120 $mT and a temperature of
approximately %$T<107 mK$, 
$T<0.1 K$
feasible in state-of-the-art experiments.   In turn, if we express the LZ velocity as $v_s= \eta v_{s,0}$, where $\eta$ is dimensionless parameter, and we take $g\mu_B \Delta b=4j$, 
the LZ sweep time is  $\Delta t_{LZ}=\frac{\hbar}{\eta j}\simeq 0.28 ns$, for $\eta\simeq 0.127$, the value of figure 4. Hence, triangulene's Haldane chain with $N=22$ units would allow for remote control between edge spins at a distance of 13 nanometers in a fraction of nanosecond.  Much faster sweeping times can be achieved with  multiple sweeps in the so-called Landau-Zener-Stuckelberg-Majorana\cite{ivakhnenko2023,rodriguez25}.

%On the other hand, the condition $g\mu_B B< j$ is required to keep the $S^z=0$ states lower in energy than the states with $S^z=\pm 1$ and avoid thermal leakage. 

%\redmark{ In the SM \cite{supp...}, we show the thermal occupation of the ground state as a function of thermal energy and the local perturbation and we discuss scenarios that optimize that occupation.} \redmark{We note that to implement full magnetization switching, the sign of the local perturbation needs to change. One possible approach is to combine a local tip-induced field with a uniform background field from a ferromagnetic substrate, separated from the spin chain by a decoupling layer. ...}

In conclusion, we have demonstrated that Haldane spin chains provide a suitable platform to achieve remote control of magnetization. The formation of a singlet between two spins located at opposite sides of the chain can be exploited to induce a local magnetization in one edge upon application of a local field on the other. We derive a simple singlet-triplet qubit  Hamiltonian that encodes the many-body information in matrix elements from Eq. \ref{eq:LSO}  that can be computed with DMRG \cite{supp}. Using our Hamiltonian, we show that the remote control to switch the magnetization in nanographene spin chains in a fraction of nanosecond are feasible with state-of-the-art.

%    x x x x x x x  x x x x x
%%^^^^00000000000000000000000ddddd

JFR  acknowledges financial support from 
%1
FCT (Grant No. PTDC/FIS-MAC/2045/2021),
%2
SNF Sinergia (Grant Pimag),
% 3
Generalitat Valenciana funding Prometeo2021/017
%4
and MFA/2022/045, and funding from MICIIN-Spain (Grant No. PID2019-109539GB-C41).
YDC acknowledges funding from FCT, QPI,  (Grant No.SFRH/BD/151311/2021).
AF acknowledges CONICET (PIP11220200100170), UNNE and the hospitality of Departamento de Física Aplicada at UA.

\appendix

%\section{Dipolar interaction}

%We now consider the effect of inter-edge dipolar interaction. We represent the dipolar operator:
%\begin{equation}
%\end{equation}

\bibliographystyle{apsrev4-2}
\bibliography{biblio}{}

\begin{thebibliography}{54}%
\makeatletter
\providecommand \@ifxundefined [1]{%
 \@ifx{#1\undefined}
}%
\providecommand \@ifnum [1]{%
 \ifnum #1\expandafter \@firstoftwo
 \else \expandafter \@secondoftwo
 \fi
}%
\providecommand \@ifx [1]{%
 \ifx #1\expandafter \@firstoftwo
 \else \expandafter \@secondoftwo
 \fi
}%
\providecommand \natexlab [1]{#1}%
\providecommand \enquote  [1]{``#1''}%
\providecommand \bibnamefont  [1]{#1}%
\providecommand \bibfnamefont [1]{#1}%
\providecommand \citenamefont [1]{#1}%
\providecommand \href@noop [0]{\@secondoftwo}%
\providecommand \href [0]{\begingroup \@sanitize@url \@href}%
\providecommand \@href[1]{\@@startlink{#1}\@@href}%
\providecommand \@@href[1]{\endgroup#1\@@endlink}%
\providecommand \@sanitize@url [0]{\catcode `\\12\catcode `\$12\catcode `\&12\catcode `\#12\catcode `\^12\catcode `\_12\catcode `\%12\relax}%
\providecommand \@@startlink[1]{}%
\providecommand \@@endlink[0]{}%
\providecommand \url  [0]{\begingroup\@sanitize@url \@url }%
\providecommand \@url [1]{\endgroup\@href {#1}{\urlprefix }}%
\providecommand \urlprefix  [0]{URL }%
\providecommand \Eprint [0]{\href }%
\providecommand \doibase [0]{https://doi.org/}%
\providecommand \selectlanguage [0]{\@gobble}%
\providecommand \bibinfo  [0]{\@secondoftwo}%
\providecommand \bibfield  [0]{\@secondoftwo}%
\providecommand \translation [1]{[#1]}%
\providecommand \BibitemOpen [0]{}%
\providecommand \bibitemStop [0]{}%
\providecommand \bibitemNoStop [0]{.\EOS\space}%
\providecommand \EOS [0]{\spacefactor3000\relax}%
\providecommand \BibitemShut  [1]{\csname bibitem#1\endcsname}%
\let\auto@bib@innerbib\@empty
%</preamble>
\bibitem [{\citenamefont {Haldane}(1983{\natexlab{a}})}]{haldane83a}%
  \BibitemOpen
  \bibfield  {author} {\bibinfo {author} {\bibfnamefont {F.}~\bibnamefont {Haldane}},\ }\href {https://doi.org/https://doi.org/10.1016/0375-9601(83)90631-X} {\bibfield  {journal} {\bibinfo  {journal} {Physics Letters A}\ }\textbf {\bibinfo {volume} {93}},\ \bibinfo {pages} {464} (\bibinfo {year} {1983}{\natexlab{a}})}\BibitemShut {NoStop}%
\bibitem [{\citenamefont {Haldane}(1983{\natexlab{b}})}]{haldane83b}%
  \BibitemOpen
  \bibfield  {author} {\bibinfo {author} {\bibfnamefont {F.~D.~M.}\ \bibnamefont {Haldane}},\ }\href {https://doi.org/10.1103/PhysRevLett.50.1153} {\bibfield  {journal} {\bibinfo  {journal} {Phys. Rev. Lett.}\ }\textbf {\bibinfo {volume} {50}},\ \bibinfo {pages} {1153} (\bibinfo {year} {1983}{\natexlab{b}})}\BibitemShut {NoStop}%
\bibitem [{\citenamefont {des Cloizeaux}\ and\ \citenamefont {Pearson}(1962)}]{Clozieaux62}%
  \BibitemOpen
  \bibfield  {author} {\bibinfo {author} {\bibfnamefont {J.}~\bibnamefont {des Cloizeaux}}\ and\ \bibinfo {author} {\bibfnamefont {J.~J.}\ \bibnamefont {Pearson}},\ }\href {https://doi.org/10.1103/PhysRev.128.2131} {\bibfield  {journal} {\bibinfo  {journal} {Phys. Rev.}\ }\textbf {\bibinfo {volume} {128}},\ \bibinfo {pages} {2131} (\bibinfo {year} {1962})}\BibitemShut {NoStop}%
\bibitem [{\citenamefont {Kennedy}(1990)}]{kennedy90}%
  \BibitemOpen
  \bibfield  {author} {\bibinfo {author} {\bibfnamefont {T.}~\bibnamefont {Kennedy}},\ }\href@noop {} {\bibfield  {journal} {\bibinfo  {journal} {Journal of Physics: Condensed Matter}\ }\textbf {\bibinfo {volume} {2}},\ \bibinfo {pages} {5737} (\bibinfo {year} {1990})}\BibitemShut {NoStop}%
\bibitem [{\citenamefont {Bose}(2003)}]{Bose2003}%
  \BibitemOpen
  \bibfield  {author} {\bibinfo {author} {\bibfnamefont {S.}~\bibnamefont {Bose}},\ }\href {https://doi.org/10.1103/PhysRevLett.91.207901} {\bibfield  {journal} {\bibinfo  {journal} {Phys. Rev. Lett.}\ }\textbf {\bibinfo {volume} {91}},\ \bibinfo {pages} {207901} (\bibinfo {year} {2003})}\BibitemShut {NoStop}%
\bibitem [{\citenamefont {Christandl}\ \emph {et~al.}(2004)\citenamefont {Christandl}, \citenamefont {Datta}, \citenamefont {Ekert},\ and\ \citenamefont {Landahl}}]{Christandl04}%
  \BibitemOpen
  \bibfield  {author} {\bibinfo {author} {\bibfnamefont {M.}~\bibnamefont {Christandl}}, \bibinfo {author} {\bibfnamefont {N.}~\bibnamefont {Datta}}, \bibinfo {author} {\bibfnamefont {A.}~\bibnamefont {Ekert}},\ and\ \bibinfo {author} {\bibfnamefont {A.~J.}\ \bibnamefont {Landahl}},\ }\href {https://doi.org/10.1103/PhysRevLett.92.187902} {\bibfield  {journal} {\bibinfo  {journal} {Phys. Rev. Lett.}\ }\textbf {\bibinfo {volume} {92}},\ \bibinfo {pages} {187902} (\bibinfo {year} {2004})}\BibitemShut {NoStop}%
\bibitem [{\citenamefont {Burgarth}\ \emph {et~al.}(2005)\citenamefont {Burgarth}, \citenamefont {Giovannetti},\ and\ \citenamefont {Bose}}]{Burgarth2005}%
  \BibitemOpen
  \bibfield  {author} {\bibinfo {author} {\bibfnamefont {D.}~\bibnamefont {Burgarth}}, \bibinfo {author} {\bibfnamefont {V.}~\bibnamefont {Giovannetti}},\ and\ \bibinfo {author} {\bibfnamefont {S.}~\bibnamefont {Bose}},\ }\href {https://doi.org/10.1088/0305-4470/38/30/013} {\bibfield  {journal} {\bibinfo  {journal} {Journal of Physics A: Mathematical and General}\ }\textbf {\bibinfo {volume} {38}},\ \bibinfo {pages} {6793} (\bibinfo {year} {2005})}\BibitemShut {NoStop}%
\bibitem [{\citenamefont {Bose}(2007)}]{bose2007}%
  \BibitemOpen
  \bibfield  {author} {\bibinfo {author} {\bibfnamefont {S.}~\bibnamefont {Bose}},\ }\href@noop {} {\bibfield  {journal} {\bibinfo  {journal} {Contemporary Physics}\ }\textbf {\bibinfo {volume} {48}},\ \bibinfo {pages} {13} (\bibinfo {year} {2007})}\BibitemShut {NoStop}%
\bibitem [{\citenamefont {Campos~Venuti}\ \emph {et~al.}(2007)\citenamefont {Campos~Venuti}, \citenamefont {Degli Esposti~Boschi},\ and\ \citenamefont {Roncaglia}}]{campos07}%
  \BibitemOpen
  \bibfield  {author} {\bibinfo {author} {\bibfnamefont {L.}~\bibnamefont {Campos~Venuti}}, \bibinfo {author} {\bibfnamefont {C.}~\bibnamefont {Degli Esposti~Boschi}},\ and\ \bibinfo {author} {\bibfnamefont {M.}~\bibnamefont {Roncaglia}},\ }\href@noop {} {\bibfield  {journal} {\bibinfo  {journal} {Physical Review Letters}\ }\textbf {\bibinfo {volume} {99}},\ \bibinfo {pages} {060401} (\bibinfo {year} {2007})}\BibitemShut {NoStop}%
\bibitem [{\citenamefont {Romero-Isart}\ \emph {et~al.}(2007)\citenamefont {Romero-Isart}, \citenamefont {Eckert},\ and\ \citenamefont {Sanpera}}]{romero07}%
  \BibitemOpen
  \bibfield  {author} {\bibinfo {author} {\bibfnamefont {O.}~\bibnamefont {Romero-Isart}}, \bibinfo {author} {\bibfnamefont {K.}~\bibnamefont {Eckert}},\ and\ \bibinfo {author} {\bibfnamefont {A.}~\bibnamefont {Sanpera}},\ }\href@noop {} {\bibfield  {journal} {\bibinfo  {journal} {Physical Review A—Atomic, Molecular, and Optical Physics}\ }\textbf {\bibinfo {volume} {75}},\ \bibinfo {pages} {050303} (\bibinfo {year} {2007})}\BibitemShut {NoStop}%
\bibitem [{\citenamefont {Zwick}\ \emph {et~al.}(2011)\citenamefont {Zwick}, \citenamefont {Alvarez}, \citenamefont {Stolze},\ and\ \citenamefont {Osenda}}]{zwick11}%
  \BibitemOpen
  \bibfield  {author} {\bibinfo {author} {\bibfnamefont {A.}~\bibnamefont {Zwick}}, \bibinfo {author} {\bibfnamefont {G.~A.}\ \bibnamefont {Alvarez}}, \bibinfo {author} {\bibfnamefont {J.}~\bibnamefont {Stolze}},\ and\ \bibinfo {author} {\bibfnamefont {O.}~\bibnamefont {Osenda}},\ }\href@noop {} {\bibfield  {journal} {\bibinfo  {journal} {Physical Review A—Atomic, Molecular, and Optical Physics}\ }\textbf {\bibinfo {volume} {84}},\ \bibinfo {pages} {022311} (\bibinfo {year} {2011})}\BibitemShut {NoStop}%
\bibitem [{\citenamefont {Nikolopoulos}\ \emph {et~al.}(2014)\citenamefont {Nikolopoulos}, \citenamefont {Jex} \emph {et~al.}}]{nikolopoulos2014}%
  \BibitemOpen
  \bibfield  {author} {\bibinfo {author} {\bibfnamefont {G.~M.}\ \bibnamefont {Nikolopoulos}}, \bibinfo {author} {\bibfnamefont {I.}~\bibnamefont {Jex}}, \emph {et~al.},\ }\href@noop {} {\emph {\bibinfo {title} {Quantum state transfer and network engineering}}}\ (\bibinfo  {publisher} {Springer},\ \bibinfo {year} {2014})\BibitemShut {NoStop}%
\bibitem [{\citenamefont {Banchi}\ \emph {et~al.}(2017)\citenamefont {Banchi}, \citenamefont {Fern{\'a}ndez-Rossier}, \citenamefont {Hirjibehedin},\ and\ \citenamefont {Bose}}]{banchi17}%
  \BibitemOpen
  \bibfield  {author} {\bibinfo {author} {\bibfnamefont {L.}~\bibnamefont {Banchi}}, \bibinfo {author} {\bibfnamefont {J.}~\bibnamefont {Fern{\'a}ndez-Rossier}}, \bibinfo {author} {\bibfnamefont {C.~F.}\ \bibnamefont {Hirjibehedin}},\ and\ \bibinfo {author} {\bibfnamefont {S.}~\bibnamefont {Bose}},\ }\href@noop {} {\bibfield  {journal} {\bibinfo  {journal} {Physical review letters}\ }\textbf {\bibinfo {volume} {118}},\ \bibinfo {pages} {147203} (\bibinfo {year} {2017})}\BibitemShut {NoStop}%
\bibitem [{\citenamefont {Bazhanov}\ \emph {et~al.}(2018)\citenamefont {Bazhanov}, \citenamefont {Sivkov},\ and\ \citenamefont {Stepanyuk}}]{bazhanov18}%
  \BibitemOpen
  \bibfield  {author} {\bibinfo {author} {\bibfnamefont {D.~I.}\ \bibnamefont {Bazhanov}}, \bibinfo {author} {\bibfnamefont {I.~N.}\ \bibnamefont {Sivkov}},\ and\ \bibinfo {author} {\bibfnamefont {V.~S.}\ \bibnamefont {Stepanyuk}},\ }\href@noop {} {\bibfield  {journal} {\bibinfo  {journal} {Scientific Reports}\ }\textbf {\bibinfo {volume} {8}},\ \bibinfo {pages} {14118} (\bibinfo {year} {2018})}\BibitemShut {NoStop}%
\bibitem [{\citenamefont {Van~Dyke}\ \emph {et~al.}(2021)\citenamefont {Van~Dyke}, \citenamefont {Kandel}, \citenamefont {Qiao}, \citenamefont {Nichol}, \citenamefont {Economou},\ and\ \citenamefont {Barnes}}]{van21}%
  \BibitemOpen
  \bibfield  {author} {\bibinfo {author} {\bibfnamefont {J.~S.}\ \bibnamefont {Van~Dyke}}, \bibinfo {author} {\bibfnamefont {Y.~P.}\ \bibnamefont {Kandel}}, \bibinfo {author} {\bibfnamefont {H.}~\bibnamefont {Qiao}}, \bibinfo {author} {\bibfnamefont {J.~M.}\ \bibnamefont {Nichol}}, \bibinfo {author} {\bibfnamefont {S.~E.}\ \bibnamefont {Economou}},\ and\ \bibinfo {author} {\bibfnamefont {E.}~\bibnamefont {Barnes}},\ }\href@noop {} {\bibfield  {journal} {\bibinfo  {journal} {Physical Review B}\ }\textbf {\bibinfo {volume} {103}},\ \bibinfo {pages} {245303} (\bibinfo {year} {2021})}\BibitemShut {NoStop}%
\bibitem [{\citenamefont {Acosta~Coden}\ \emph {et~al.}(2024)\citenamefont {Acosta~Coden}, \citenamefont {Osenda},\ and\ \citenamefont {Ferrón}}]{Coden2024}%
  \BibitemOpen
  \bibfield  {author} {\bibinfo {author} {\bibfnamefont {D.~S.}\ \bibnamefont {Acosta~Coden}}, \bibinfo {author} {\bibfnamefont {O.}~\bibnamefont {Osenda}},\ and\ \bibinfo {author} {\bibfnamefont {A.}~\bibnamefont {Ferrón}},\ }\href {https://doi.org/10.1088/1361-6455/ad9a30} {\bibfield  {journal} {\bibinfo  {journal} {Journal of Physics B: Atomic, Molecular and Optical Physics}\ }\textbf {\bibinfo {volume} {58}},\ \bibinfo {pages} {015504} (\bibinfo {year} {2024})}\BibitemShut {NoStop}%
\bibitem [{\citenamefont {Hirjibehedin}\ \emph {et~al.}(2006)\citenamefont {Hirjibehedin}, \citenamefont {Lutz},\ and\ \citenamefont {Heinrich}}]{hirjibehedin06}%
  \BibitemOpen
  \bibfield  {author} {\bibinfo {author} {\bibfnamefont {C.~F.}\ \bibnamefont {Hirjibehedin}}, \bibinfo {author} {\bibfnamefont {C.~P.}\ \bibnamefont {Lutz}},\ and\ \bibinfo {author} {\bibfnamefont {A.~J.}\ \bibnamefont {Heinrich}},\ }\href@noop {} {\bibfield  {journal} {\bibinfo  {journal} {Science}\ }\textbf {\bibinfo {volume} {312}},\ \bibinfo {pages} {1021} (\bibinfo {year} {2006})}\BibitemShut {NoStop}%
\bibitem [{\citenamefont {Khajetoorians}\ \emph {et~al.}(2011)\citenamefont {Khajetoorians}, \citenamefont {Wiebe}, \citenamefont {Chilian},\ and\ \citenamefont {Wiesendanger}}]{khajetoorians11}%
  \BibitemOpen
  \bibfield  {author} {\bibinfo {author} {\bibfnamefont {A.~A.}\ \bibnamefont {Khajetoorians}}, \bibinfo {author} {\bibfnamefont {J.}~\bibnamefont {Wiebe}}, \bibinfo {author} {\bibfnamefont {B.}~\bibnamefont {Chilian}},\ and\ \bibinfo {author} {\bibfnamefont {R.}~\bibnamefont {Wiesendanger}},\ }\href@noop {} {\bibfield  {journal} {\bibinfo  {journal} {Science}\ }\textbf {\bibinfo {volume} {332}},\ \bibinfo {pages} {1062} (\bibinfo {year} {2011})}\BibitemShut {NoStop}%
\bibitem [{\citenamefont {Choi}\ \emph {et~al.}(2019)\citenamefont {Choi}, \citenamefont {Lorente}, \citenamefont {Wiebe}, \citenamefont {Von~Bergmann}, \citenamefont {Otte},\ and\ \citenamefont {Heinrich}}]{choi19}%
  \BibitemOpen
  \bibfield  {author} {\bibinfo {author} {\bibfnamefont {D.-J.}\ \bibnamefont {Choi}}, \bibinfo {author} {\bibfnamefont {N.}~\bibnamefont {Lorente}}, \bibinfo {author} {\bibfnamefont {J.}~\bibnamefont {Wiebe}}, \bibinfo {author} {\bibfnamefont {K.}~\bibnamefont {Von~Bergmann}}, \bibinfo {author} {\bibfnamefont {A.~F.}\ \bibnamefont {Otte}},\ and\ \bibinfo {author} {\bibfnamefont {A.~J.}\ \bibnamefont {Heinrich}},\ }\href@noop {} {\bibfield  {journal} {\bibinfo  {journal} {Reviews of Modern Physics}\ }\textbf {\bibinfo {volume} {91}},\ \bibinfo {pages} {041001} (\bibinfo {year} {2019})}\BibitemShut {NoStop}%
\bibitem [{\citenamefont {Mishra}\ \emph {et~al.}(2021)\citenamefont {Mishra}, \citenamefont {Catarina}, \citenamefont {Wu}, \citenamefont {Ortiz}, \citenamefont {Jacob}, \citenamefont {Eimre}, \citenamefont {Ma}, \citenamefont {Pignedoli}, \citenamefont {Feng}, \citenamefont {Ruffieux}, \citenamefont {Fernandez-Rossier},\ and\ \citenamefont {Fasel}}]{Mishra2021}%
  \BibitemOpen
  \bibfield  {author} {\bibinfo {author} {\bibfnamefont {S.}~\bibnamefont {Mishra}}, \bibinfo {author} {\bibfnamefont {G.}~\bibnamefont {Catarina}}, \bibinfo {author} {\bibfnamefont {F.}~\bibnamefont {Wu}}, \bibinfo {author} {\bibfnamefont {R.}~\bibnamefont {Ortiz}}, \bibinfo {author} {\bibfnamefont {D.}~\bibnamefont {Jacob}}, \bibinfo {author} {\bibfnamefont {K.}~\bibnamefont {Eimre}}, \bibinfo {author} {\bibfnamefont {J.}~\bibnamefont {Ma}}, \bibinfo {author} {\bibfnamefont {C.~A.}\ \bibnamefont {Pignedoli}}, \bibinfo {author} {\bibfnamefont {X.}~\bibnamefont {Feng}}, \bibinfo {author} {\bibfnamefont {P.}~\bibnamefont {Ruffieux}}, \bibinfo {author} {\bibfnamefont {J.}~\bibnamefont {Fernandez-Rossier}},\ and\ \bibinfo {author} {\bibfnamefont {R.}~\bibnamefont {Fasel}},\ }\href@noop {} {\bibfield  {journal} {\bibinfo  {journal} {Nature}\ }\textbf {\bibinfo {volume} {598}},\ \bibinfo {pages} {287} (\bibinfo {year} {2021})}\BibitemShut {NoStop}%
\bibitem [{\citenamefont {Zhao}\ \emph {et~al.}(2024)\citenamefont {Zhao}, \citenamefont {Catarina}, \citenamefont {Zhang}, \citenamefont {Henriques}, \citenamefont {Yang}, \citenamefont {Ma}, \citenamefont {Feng}, \citenamefont {Gr{\"o}ning}, \citenamefont {Ruffieux}, \citenamefont {Fern{\'a}ndez-Rossier} \emph {et~al.}}]{zhao24}%
  \BibitemOpen
  \bibfield  {author} {\bibinfo {author} {\bibfnamefont {C.}~\bibnamefont {Zhao}}, \bibinfo {author} {\bibfnamefont {G.}~\bibnamefont {Catarina}}, \bibinfo {author} {\bibfnamefont {J.-J.}\ \bibnamefont {Zhang}}, \bibinfo {author} {\bibfnamefont {J.~C.}\ \bibnamefont {Henriques}}, \bibinfo {author} {\bibfnamefont {L.}~\bibnamefont {Yang}}, \bibinfo {author} {\bibfnamefont {J.}~\bibnamefont {Ma}}, \bibinfo {author} {\bibfnamefont {X.}~\bibnamefont {Feng}}, \bibinfo {author} {\bibfnamefont {O.}~\bibnamefont {Gr{\"o}ning}}, \bibinfo {author} {\bibfnamefont {P.}~\bibnamefont {Ruffieux}}, \bibinfo {author} {\bibfnamefont {J.}~\bibnamefont {Fern{\'a}ndez-Rossier}}, \emph {et~al.},\ }\href@noop {} {\bibfield  {journal} {\bibinfo  {journal} {Nature Nanotechnology}\ ,\ \bibinfo {pages} {1}} (\bibinfo {year} {2024})}\BibitemShut {NoStop}%
\bibitem [{\citenamefont {Zhao}\ \emph {et~al.}(2025)\citenamefont {Zhao}, \citenamefont {Yang}, \citenamefont {Henriques}, \citenamefont {Ferri-Cort{\'e}s}, \citenamefont {Catarina}, \citenamefont {Pignedoli}, \citenamefont {Ma}, \citenamefont {Feng}, \citenamefont {Ruffieux}, \citenamefont {Fern{\'a}ndez-Rossier} \emph {et~al.}}]{zhao25}%
  \BibitemOpen
  \bibfield  {author} {\bibinfo {author} {\bibfnamefont {C.}~\bibnamefont {Zhao}}, \bibinfo {author} {\bibfnamefont {L.}~\bibnamefont {Yang}}, \bibinfo {author} {\bibfnamefont {J.~C.}\ \bibnamefont {Henriques}}, \bibinfo {author} {\bibfnamefont {M.}~\bibnamefont {Ferri-Cort{\'e}s}}, \bibinfo {author} {\bibfnamefont {G.}~\bibnamefont {Catarina}}, \bibinfo {author} {\bibfnamefont {C.~A.}\ \bibnamefont {Pignedoli}}, \bibinfo {author} {\bibfnamefont {J.}~\bibnamefont {Ma}}, \bibinfo {author} {\bibfnamefont {X.}~\bibnamefont {Feng}}, \bibinfo {author} {\bibfnamefont {P.}~\bibnamefont {Ruffieux}}, \bibinfo {author} {\bibfnamefont {J.}~\bibnamefont {Fern{\'a}ndez-Rossier}}, \emph {et~al.},\ }\href@noop {} {\bibfield  {journal} {\bibinfo  {journal} {Nature Materials}\ ,\ \bibinfo {pages} {1}} (\bibinfo {year} {2025})}\BibitemShut {NoStop}%
\bibitem [{\citenamefont {van Diepen}\ \emph {et~al.}(2021)\citenamefont {van Diepen}, \citenamefont {Hsiao}, \citenamefont {Mukhopadhyay}, \citenamefont {Reichl}, \citenamefont {Wegscheider},\ and\ \citenamefont {Vandersypen}}]{vandiepen21}%
  \BibitemOpen
  \bibfield  {author} {\bibinfo {author} {\bibfnamefont {C.~J.}\ \bibnamefont {van Diepen}}, \bibinfo {author} {\bibfnamefont {T.-K.}\ \bibnamefont {Hsiao}}, \bibinfo {author} {\bibfnamefont {U.}~\bibnamefont {Mukhopadhyay}}, \bibinfo {author} {\bibfnamefont {C.}~\bibnamefont {Reichl}}, \bibinfo {author} {\bibfnamefont {W.}~\bibnamefont {Wegscheider}},\ and\ \bibinfo {author} {\bibfnamefont {L.~M.}\ \bibnamefont {Vandersypen}},\ }\href@noop {} {\bibfield  {journal} {\bibinfo  {journal} {Physical Review X}\ }\textbf {\bibinfo {volume} {11}},\ \bibinfo {pages} {041025} (\bibinfo {year} {2021})}\BibitemShut {NoStop}%
\bibitem [{\citenamefont {Donnelly}\ \emph {et~al.}(2024)\citenamefont {Donnelly}, \citenamefont {Rowlands}, \citenamefont {Kranz}, \citenamefont {Hsueh}, \citenamefont {Chung}, \citenamefont {Timofeev}, \citenamefont {Geng}, \citenamefont {Singh-Gregory}, \citenamefont {Gorman}, \citenamefont {Keizer} \emph {et~al.}}]{donnelly24}%
  \BibitemOpen
  \bibfield  {author} {\bibinfo {author} {\bibfnamefont {M.}~\bibnamefont {Donnelly}}, \bibinfo {author} {\bibfnamefont {J.}~\bibnamefont {Rowlands}}, \bibinfo {author} {\bibfnamefont {L.}~\bibnamefont {Kranz}}, \bibinfo {author} {\bibfnamefont {Y.}~\bibnamefont {Hsueh}}, \bibinfo {author} {\bibfnamefont {Y.}~\bibnamefont {Chung}}, \bibinfo {author} {\bibfnamefont {A.}~\bibnamefont {Timofeev}}, \bibinfo {author} {\bibfnamefont {H.}~\bibnamefont {Geng}}, \bibinfo {author} {\bibfnamefont {P.}~\bibnamefont {Singh-Gregory}}, \bibinfo {author} {\bibfnamefont {S.}~\bibnamefont {Gorman}}, \bibinfo {author} {\bibfnamefont {J.}~\bibnamefont {Keizer}}, \emph {et~al.},\ }\href@noop {} {\bibfield  {journal} {\bibinfo  {journal} {arXiv preprint arXiv:2405.03763}\ } (\bibinfo {year} {2024})}\BibitemShut {NoStop}%
\bibitem [{\citenamefont {Murmann}\ \emph {et~al.}(2015)\citenamefont {Murmann}, \citenamefont {Deuretzbacher}, \citenamefont {Z\"urn}, \citenamefont {Bjerlin}, \citenamefont {Reimann}, \citenamefont {Santos}, \citenamefont {Lompe},\ and\ \citenamefont {Jochim}}]{Murmann2015}%
  \BibitemOpen
  \bibfield  {author} {\bibinfo {author} {\bibfnamefont {S.}~\bibnamefont {Murmann}}, \bibinfo {author} {\bibfnamefont {F.}~\bibnamefont {Deuretzbacher}}, \bibinfo {author} {\bibfnamefont {G.}~\bibnamefont {Z\"urn}}, \bibinfo {author} {\bibfnamefont {J.}~\bibnamefont {Bjerlin}}, \bibinfo {author} {\bibfnamefont {S.~M.}\ \bibnamefont {Reimann}}, \bibinfo {author} {\bibfnamefont {L.}~\bibnamefont {Santos}}, \bibinfo {author} {\bibfnamefont {T.}~\bibnamefont {Lompe}},\ and\ \bibinfo {author} {\bibfnamefont {S.}~\bibnamefont {Jochim}},\ }\href {https://doi.org/10.1103/PhysRevLett.115.215301} {\bibfield  {journal} {\bibinfo  {journal} {Phys. Rev. Lett.}\ }\textbf {\bibinfo {volume} {115}},\ \bibinfo {pages} {215301} (\bibinfo {year} {2015})}\BibitemShut {NoStop}%
\bibitem [{\citenamefont {Senko}\ \emph {et~al.}(2015)\citenamefont {Senko}, \citenamefont {Richerme}, \citenamefont {Smith}, \citenamefont {Lee}, \citenamefont {Cohen}, \citenamefont {Retzker},\ and\ \citenamefont {Monroe}}]{senko15}%
  \BibitemOpen
  \bibfield  {author} {\bibinfo {author} {\bibfnamefont {C.}~\bibnamefont {Senko}}, \bibinfo {author} {\bibfnamefont {P.}~\bibnamefont {Richerme}}, \bibinfo {author} {\bibfnamefont {J.}~\bibnamefont {Smith}}, \bibinfo {author} {\bibfnamefont {A.}~\bibnamefont {Lee}}, \bibinfo {author} {\bibfnamefont {I.}~\bibnamefont {Cohen}}, \bibinfo {author} {\bibfnamefont {A.}~\bibnamefont {Retzker}},\ and\ \bibinfo {author} {\bibfnamefont {C.}~\bibnamefont {Monroe}},\ }\href@noop {} {\bibfield  {journal} {\bibinfo  {journal} {Physical Review X}\ }\textbf {\bibinfo {volume} {5}},\ \bibinfo {pages} {021026} (\bibinfo {year} {2015})}\BibitemShut {NoStop}%
\bibitem [{\citenamefont {Barredo}\ \emph {et~al.}(2015)\citenamefont {Barredo}, \citenamefont {Labuhn}, \citenamefont {Ravets}, \citenamefont {Lahaye}, \citenamefont {Browaeys},\ and\ \citenamefont {Adams}}]{barredo15}%
  \BibitemOpen
  \bibfield  {author} {\bibinfo {author} {\bibfnamefont {D.}~\bibnamefont {Barredo}}, \bibinfo {author} {\bibfnamefont {H.}~\bibnamefont {Labuhn}}, \bibinfo {author} {\bibfnamefont {S.}~\bibnamefont {Ravets}}, \bibinfo {author} {\bibfnamefont {T.}~\bibnamefont {Lahaye}}, \bibinfo {author} {\bibfnamefont {A.}~\bibnamefont {Browaeys}},\ and\ \bibinfo {author} {\bibfnamefont {C.~S.}\ \bibnamefont {Adams}},\ }\href@noop {} {\bibfield  {journal} {\bibinfo  {journal} {Physical review letters}\ }\textbf {\bibinfo {volume} {114}},\ \bibinfo {pages} {113002} (\bibinfo {year} {2015})}\BibitemShut {NoStop}%
\bibitem [{\citenamefont {Kandel}\ \emph {et~al.}(2021)\citenamefont {Kandel}, \citenamefont {Qiao}, \citenamefont {Fallahi}, \citenamefont {Gardner}, \citenamefont {Manfra},\ and\ \citenamefont {Nichol}}]{Kandel2021}%
  \BibitemOpen
  \bibfield  {author} {\bibinfo {author} {\bibfnamefont {P.~K.}\ \bibnamefont {Kandel}}, \bibinfo {author} {\bibfnamefont {H.}~\bibnamefont {Qiao}}, \bibinfo {author} {\bibfnamefont {S.}~\bibnamefont {Fallahi}}, \bibinfo {author} {\bibfnamefont {G.~C.}\ \bibnamefont {Gardner}}, \bibinfo {author} {\bibfnamefont {M.~J.}\ \bibnamefont {Manfra}},\ and\ \bibinfo {author} {\bibfnamefont {J.~M.}\ \bibnamefont {Nichol}},\ }\href {https://doi.org/10.1038/s41467-021-22416-5} {\bibfield  {journal} {\bibinfo  {journal} {Nature Communications}\ }\textbf {\bibinfo {volume} {12}},\ \bibinfo {pages} {2156} (\bibinfo {year} {2021})}\BibitemShut {NoStop}%
\bibitem [{\citenamefont {Wang}\ \emph {et~al.}(2023)\citenamefont {Wang}, \citenamefont {Chen}, \citenamefont {Bui}, \citenamefont {Wolf}, \citenamefont {Haze}, \citenamefont {Mier}, \citenamefont {Kim}, \citenamefont {Choi}, \citenamefont {Lutz}, \citenamefont {Bae} \emph {et~al.}}]{Wang2023}%
  \BibitemOpen
  \bibfield  {author} {\bibinfo {author} {\bibfnamefont {Y.}~\bibnamefont {Wang}}, \bibinfo {author} {\bibfnamefont {Y.}~\bibnamefont {Chen}}, \bibinfo {author} {\bibfnamefont {H.~T.}\ \bibnamefont {Bui}}, \bibinfo {author} {\bibfnamefont {C.}~\bibnamefont {Wolf}}, \bibinfo {author} {\bibfnamefont {M.}~\bibnamefont {Haze}}, \bibinfo {author} {\bibfnamefont {C.}~\bibnamefont {Mier}}, \bibinfo {author} {\bibfnamefont {J.}~\bibnamefont {Kim}}, \bibinfo {author} {\bibfnamefont {D.-J.}\ \bibnamefont {Choi}}, \bibinfo {author} {\bibfnamefont {C.~P.}\ \bibnamefont {Lutz}}, \bibinfo {author} {\bibfnamefont {Y.}~\bibnamefont {Bae}}, \emph {et~al.},\ }\href@noop {} {\bibfield  {journal} {\bibinfo  {journal} {Science}\ }\textbf {\bibinfo {volume} {382}},\ \bibinfo {pages} {87} (\bibinfo {year} {2023})}\BibitemShut {NoStop}%
\bibitem [{\citenamefont {Hida}(1992)}]{hida92}%
  \BibitemOpen
  \bibfield  {author} {\bibinfo {author} {\bibfnamefont {K.}~\bibnamefont {Hida}},\ }\href@noop {} {\bibfield  {journal} {\bibinfo  {journal} {Physical Review B}\ }\textbf {\bibinfo {volume} {45}},\ \bibinfo {pages} {2207} (\bibinfo {year} {1992})}\BibitemShut {NoStop}%
\bibitem [{\citenamefont {Wang}\ \emph {et~al.}(2024)\citenamefont {Wang}, \citenamefont {Fan}, \citenamefont {Chen}, \citenamefont {Jiang}, \citenamefont {Gao}, \citenamefont {Lado},\ and\ \citenamefont {Yang}}]{wang24}%
  \BibitemOpen
  \bibfield  {author} {\bibinfo {author} {\bibfnamefont {H.}~\bibnamefont {Wang}}, \bibinfo {author} {\bibfnamefont {P.}~\bibnamefont {Fan}}, \bibinfo {author} {\bibfnamefont {J.}~\bibnamefont {Chen}}, \bibinfo {author} {\bibfnamefont {L.}~\bibnamefont {Jiang}}, \bibinfo {author} {\bibfnamefont {H.-J.}\ \bibnamefont {Gao}}, \bibinfo {author} {\bibfnamefont {J.~L.}\ \bibnamefont {Lado}},\ and\ \bibinfo {author} {\bibfnamefont {K.}~\bibnamefont {Yang}},\ }\href@noop {} {\bibfield  {journal} {\bibinfo  {journal} {Nature Nanotechnology}\ ,\ \bibinfo {pages} {1}} (\bibinfo {year} {2024})}\BibitemShut {NoStop}%
\bibitem [{\citenamefont {Sompet}\ \emph {et~al.}(2022)\citenamefont {Sompet}, \citenamefont {Hirthe}, \citenamefont {Bourgund}, \citenamefont {Chalopin}, \citenamefont {Bibo}, \citenamefont {Koepsell}, \citenamefont {Bojovi{\'c}}, \citenamefont {Verresen}, \citenamefont {Pollmann}, \citenamefont {Salomon} \emph {et~al.}}]{sompet22}%
  \BibitemOpen
  \bibfield  {author} {\bibinfo {author} {\bibfnamefont {P.}~\bibnamefont {Sompet}}, \bibinfo {author} {\bibfnamefont {S.}~\bibnamefont {Hirthe}}, \bibinfo {author} {\bibfnamefont {D.}~\bibnamefont {Bourgund}}, \bibinfo {author} {\bibfnamefont {T.}~\bibnamefont {Chalopin}}, \bibinfo {author} {\bibfnamefont {J.}~\bibnamefont {Bibo}}, \bibinfo {author} {\bibfnamefont {J.}~\bibnamefont {Koepsell}}, \bibinfo {author} {\bibfnamefont {P.}~\bibnamefont {Bojovi{\'c}}}, \bibinfo {author} {\bibfnamefont {R.}~\bibnamefont {Verresen}}, \bibinfo {author} {\bibfnamefont {F.}~\bibnamefont {Pollmann}}, \bibinfo {author} {\bibfnamefont {G.}~\bibnamefont {Salomon}}, \emph {et~al.},\ }\href@noop {} {\bibfield  {journal} {\bibinfo  {journal} {Nature}\ }\textbf {\bibinfo {volume} {606}},\ \bibinfo {pages} {484} (\bibinfo {year} {2022})}\BibitemShut {NoStop}%
\bibitem [{\citenamefont {Munia}\ \emph {et~al.}(2024)\citenamefont {Munia}, \citenamefont {Monir}, \citenamefont {Osika}, \citenamefont {Simmons},\ and\ \citenamefont {Rahman}}]{munia24}%
  \BibitemOpen
  \bibfield  {author} {\bibinfo {author} {\bibfnamefont {M.~M.}\ \bibnamefont {Munia}}, \bibinfo {author} {\bibfnamefont {S.}~\bibnamefont {Monir}}, \bibinfo {author} {\bibfnamefont {E.~N.}\ \bibnamefont {Osika}}, \bibinfo {author} {\bibfnamefont {M.~Y.}\ \bibnamefont {Simmons}},\ and\ \bibinfo {author} {\bibfnamefont {R.}~\bibnamefont {Rahman}},\ }\href@noop {} {\bibfield  {journal} {\bibinfo  {journal} {Physical Review Applied}\ }\textbf {\bibinfo {volume} {21}},\ \bibinfo {pages} {014038} (\bibinfo {year} {2024})}\BibitemShut {NoStop}%
\bibitem [{\citenamefont {De~L{\'e}s{\'e}leuc}\ \emph {et~al.}(2019)\citenamefont {De~L{\'e}s{\'e}leuc}, \citenamefont {Lienhard}, \citenamefont {Scholl}, \citenamefont {Barredo}, \citenamefont {Weber}, \citenamefont {Lang}, \citenamefont {B{\"u}chler}, \citenamefont {Lahaye},\ and\ \citenamefont {Browaeys}}]{de19}%
  \BibitemOpen
  \bibfield  {author} {\bibinfo {author} {\bibfnamefont {S.}~\bibnamefont {De~L{\'e}s{\'e}leuc}}, \bibinfo {author} {\bibfnamefont {V.}~\bibnamefont {Lienhard}}, \bibinfo {author} {\bibfnamefont {P.}~\bibnamefont {Scholl}}, \bibinfo {author} {\bibfnamefont {D.}~\bibnamefont {Barredo}}, \bibinfo {author} {\bibfnamefont {S.}~\bibnamefont {Weber}}, \bibinfo {author} {\bibfnamefont {N.}~\bibnamefont {Lang}}, \bibinfo {author} {\bibfnamefont {H.~P.}\ \bibnamefont {B{\"u}chler}}, \bibinfo {author} {\bibfnamefont {T.}~\bibnamefont {Lahaye}},\ and\ \bibinfo {author} {\bibfnamefont {A.}~\bibnamefont {Browaeys}},\ }\href@noop {} {\bibfield  {journal} {\bibinfo  {journal} {Science}\ }\textbf {\bibinfo {volume} {365}},\ \bibinfo {pages} {775} (\bibinfo {year} {2019})}\BibitemShut {NoStop}%
\bibitem [{\citenamefont {M{\"o}gerle}\ \emph {et~al.}(2024)\citenamefont {M{\"o}gerle}, \citenamefont {Brechtelsbauer}, \citenamefont {Gea-Caballero}, \citenamefont {Prior}, \citenamefont {Emperauger}, \citenamefont {Bornet}, \citenamefont {Chen}, \citenamefont {Lahaye}, \citenamefont {Browaeys},\ and\ \citenamefont {B{\"u}chler}}]{mogerle24}%
  \BibitemOpen
  \bibfield  {author} {\bibinfo {author} {\bibfnamefont {J.}~\bibnamefont {M{\"o}gerle}}, \bibinfo {author} {\bibfnamefont {K.}~\bibnamefont {Brechtelsbauer}}, \bibinfo {author} {\bibfnamefont {A.}~\bibnamefont {Gea-Caballero}}, \bibinfo {author} {\bibfnamefont {J.}~\bibnamefont {Prior}}, \bibinfo {author} {\bibfnamefont {G.}~\bibnamefont {Emperauger}}, \bibinfo {author} {\bibfnamefont {G.}~\bibnamefont {Bornet}}, \bibinfo {author} {\bibfnamefont {C.}~\bibnamefont {Chen}}, \bibinfo {author} {\bibfnamefont {T.}~\bibnamefont {Lahaye}}, \bibinfo {author} {\bibfnamefont {A.}~\bibnamefont {Browaeys}},\ and\ \bibinfo {author} {\bibfnamefont {H.}~\bibnamefont {B{\"u}chler}},\ }\href@noop {} {\bibfield  {journal} {\bibinfo  {journal} {arXiv preprint arXiv:2410.21424}\ } (\bibinfo {year} {2024})}\BibitemShut {NoStop}%
\bibitem [{\citenamefont {Jaworowski}\ \emph {et~al.}(2017)\citenamefont {Jaworowski}, \citenamefont {Rogers}, \citenamefont {Grabowski},\ and\ \citenamefont {Hawrylak}}]{jaworowski17}%
  \BibitemOpen
  \bibfield  {author} {\bibinfo {author} {\bibfnamefont {B.}~\bibnamefont {Jaworowski}}, \bibinfo {author} {\bibfnamefont {N.}~\bibnamefont {Rogers}}, \bibinfo {author} {\bibfnamefont {M.}~\bibnamefont {Grabowski}},\ and\ \bibinfo {author} {\bibfnamefont {P.}~\bibnamefont {Hawrylak}},\ }\href@noop {} {\bibfield  {journal} {\bibinfo  {journal} {Scientific Reports}\ }\textbf {\bibinfo {volume} {7}},\ \bibinfo {pages} {5529} (\bibinfo {year} {2017})}\BibitemShut {NoStop}%
\bibitem [{\citenamefont {Baran}\ and\ \citenamefont {Paaske}(2024)}]{baran24}%
  \BibitemOpen
  \bibfield  {author} {\bibinfo {author} {\bibfnamefont {V.~V.}\ \bibnamefont {Baran}}\ and\ \bibinfo {author} {\bibfnamefont {J.}~\bibnamefont {Paaske}},\ }\href@noop {} {\bibfield  {journal} {\bibinfo  {journal} {Physical Review B}\ }\textbf {\bibinfo {volume} {110}},\ \bibinfo {pages} {064503} (\bibinfo {year} {2024})}\BibitemShut {NoStop}%
\bibitem [{\citenamefont {Affleck}\ \emph {et~al.}(1987)\citenamefont {Affleck}, \citenamefont {Kennedy}, \citenamefont {Lieb},\ and\ \citenamefont {Tasaki}}]{Affleck87}%
  \BibitemOpen
  \bibfield  {author} {\bibinfo {author} {\bibfnamefont {I.}~\bibnamefont {Affleck}}, \bibinfo {author} {\bibfnamefont {T.}~\bibnamefont {Kennedy}}, \bibinfo {author} {\bibfnamefont {E.~H.}\ \bibnamefont {Lieb}},\ and\ \bibinfo {author} {\bibfnamefont {H.}~\bibnamefont {Tasaki}},\ }\href {https://doi.org/10.1103/PhysRevLett.59.799} {\bibfield  {journal} {\bibinfo  {journal} {Phys. Rev. Lett.}\ }\textbf {\bibinfo {volume} {59}},\ \bibinfo {pages} {799} (\bibinfo {year} {1987})}\BibitemShut {NoStop}%
\bibitem [{\citenamefont {Diederix}\ \emph {et~al.}(1979)\citenamefont {Diederix}, \citenamefont {Bl{\"o}te}, \citenamefont {Groen}, \citenamefont {Klaassen},\ and\ \citenamefont {Poulis}}]{diederix79}%
  \BibitemOpen
  \bibfield  {author} {\bibinfo {author} {\bibfnamefont {K.}~\bibnamefont {Diederix}}, \bibinfo {author} {\bibfnamefont {H.}~\bibnamefont {Bl{\"o}te}}, \bibinfo {author} {\bibfnamefont {J.}~\bibnamefont {Groen}}, \bibinfo {author} {\bibfnamefont {T.}~\bibnamefont {Klaassen}},\ and\ \bibinfo {author} {\bibfnamefont {N.}~\bibnamefont {Poulis}},\ }\href@noop {} {\bibfield  {journal} {\bibinfo  {journal} {Physical Review B}\ }\textbf {\bibinfo {volume} {19}},\ \bibinfo {pages} {420} (\bibinfo {year} {1979})}\BibitemShut {NoStop}%
\bibitem [{\citenamefont {Bonner}\ \emph {et~al.}(1983)\citenamefont {Bonner}, \citenamefont {Friedberg}, \citenamefont {Kobayashi}, \citenamefont {Meier},\ and\ \citenamefont {Bl{\"o}te}}]{bonner83}%
  \BibitemOpen
  \bibfield  {author} {\bibinfo {author} {\bibfnamefont {J.~C.}\ \bibnamefont {Bonner}}, \bibinfo {author} {\bibfnamefont {S.~A.}\ \bibnamefont {Friedberg}}, \bibinfo {author} {\bibfnamefont {H.}~\bibnamefont {Kobayashi}}, \bibinfo {author} {\bibfnamefont {D.~L.}\ \bibnamefont {Meier}},\ and\ \bibinfo {author} {\bibfnamefont {H.~W.}\ \bibnamefont {Bl{\"o}te}},\ }\href@noop {} {\bibfield  {journal} {\bibinfo  {journal} {Physical Review B}\ }\textbf {\bibinfo {volume} {27}},\ \bibinfo {pages} {248} (\bibinfo {year} {1983})}\BibitemShut {NoStop}%
\bibitem [{\citenamefont {Yang}\ \emph {et~al.}(2019)\citenamefont {Yang}, \citenamefont {Paul}, \citenamefont {Natterer}, \citenamefont {Lado}, \citenamefont {Bae}, \citenamefont {Willke}, \citenamefont {Choi}, \citenamefont {Ferr{\'o}n}, \citenamefont {Fern{\'a}ndez-Rossier}, \citenamefont {Heinrich} \emph {et~al.}}]{yang19}%
  \BibitemOpen
  \bibfield  {author} {\bibinfo {author} {\bibfnamefont {K.}~\bibnamefont {Yang}}, \bibinfo {author} {\bibfnamefont {W.}~\bibnamefont {Paul}}, \bibinfo {author} {\bibfnamefont {F.~D.}\ \bibnamefont {Natterer}}, \bibinfo {author} {\bibfnamefont {J.~L.}\ \bibnamefont {Lado}}, \bibinfo {author} {\bibfnamefont {Y.}~\bibnamefont {Bae}}, \bibinfo {author} {\bibfnamefont {P.}~\bibnamefont {Willke}}, \bibinfo {author} {\bibfnamefont {T.}~\bibnamefont {Choi}}, \bibinfo {author} {\bibfnamefont {A.}~\bibnamefont {Ferr{\'o}n}}, \bibinfo {author} {\bibfnamefont {J.}~\bibnamefont {Fern{\'a}ndez-Rossier}}, \bibinfo {author} {\bibfnamefont {A.~J.}\ \bibnamefont {Heinrich}}, \emph {et~al.},\ }\href@noop {} {\bibfield  {journal} {\bibinfo  {journal} {Physical review letters}\ }\textbf {\bibinfo {volume} {122}},\ \bibinfo {pages} {227203} (\bibinfo {year} {2019})}\BibitemShut {NoStop}%
\bibitem [{\citenamefont {Phark}\ \emph {et~al.}(2023)\citenamefont {Phark}, \citenamefont {Bui}, \citenamefont {Ferr{\'o}n}, \citenamefont {Fern{\'a}ndez-Rossier}, \citenamefont {Reina-G{\'a}lvez}, \citenamefont {Wolf}, \citenamefont {Wang}, \citenamefont {Yang}, \citenamefont {Heinrich},\ and\ \citenamefont {Lutz}}]{phark23}%
  \BibitemOpen
  \bibfield  {author} {\bibinfo {author} {\bibfnamefont {S.-h.}\ \bibnamefont {Phark}}, \bibinfo {author} {\bibfnamefont {H.~T.}\ \bibnamefont {Bui}}, \bibinfo {author} {\bibfnamefont {A.}~\bibnamefont {Ferr{\'o}n}}, \bibinfo {author} {\bibfnamefont {J.}~\bibnamefont {Fern{\'a}ndez-Rossier}}, \bibinfo {author} {\bibfnamefont {J.}~\bibnamefont {Reina-G{\'a}lvez}}, \bibinfo {author} {\bibfnamefont {C.}~\bibnamefont {Wolf}}, \bibinfo {author} {\bibfnamefont {Y.}~\bibnamefont {Wang}}, \bibinfo {author} {\bibfnamefont {K.}~\bibnamefont {Yang}}, \bibinfo {author} {\bibfnamefont {A.~J.}\ \bibnamefont {Heinrich}},\ and\ \bibinfo {author} {\bibfnamefont {C.~P.}\ \bibnamefont {Lutz}},\ }\href@noop {} {\bibfield  {journal} {\bibinfo  {journal} {Advanced Science}\ }\textbf {\bibinfo {volume} {10}},\ \bibinfo {pages} {2302033} (\bibinfo {year} {2023})}\BibitemShut {NoStop}%
\bibitem [{\citenamefont {Petta}\ \emph {et~al.}(2005)\citenamefont {Petta}, \citenamefont {Johnson}, \citenamefont {Taylor}, \citenamefont {Laird}, \citenamefont {Yacoby}, \citenamefont {Lukin}, \citenamefont {Marcus}, \citenamefont {Hanson},\ and\ \citenamefont {Gossard}}]{petta05}%
  \BibitemOpen
  \bibfield  {author} {\bibinfo {author} {\bibfnamefont {J.~R.}\ \bibnamefont {Petta}}, \bibinfo {author} {\bibfnamefont {A.~C.}\ \bibnamefont {Johnson}}, \bibinfo {author} {\bibfnamefont {J.~M.}\ \bibnamefont {Taylor}}, \bibinfo {author} {\bibfnamefont {E.~A.}\ \bibnamefont {Laird}}, \bibinfo {author} {\bibfnamefont {A.}~\bibnamefont {Yacoby}}, \bibinfo {author} {\bibfnamefont {M.~D.}\ \bibnamefont {Lukin}}, \bibinfo {author} {\bibfnamefont {C.~M.}\ \bibnamefont {Marcus}}, \bibinfo {author} {\bibfnamefont {M.~P.}\ \bibnamefont {Hanson}},\ and\ \bibinfo {author} {\bibfnamefont {A.~C.}\ \bibnamefont {Gossard}},\ }\href@noop {} {\bibfield  {journal} {\bibinfo  {journal} {Science}\ }\textbf {\bibinfo {volume} {309}},\ \bibinfo {pages} {2180} (\bibinfo {year} {2005})}\BibitemShut {NoStop}%
\bibitem [{\citenamefont {Ortiz}\ \emph {et~al.}(2018)\citenamefont {Ortiz}, \citenamefont {Garc{\'\i}a-Mart{\'\i}nez}, \citenamefont {Lado},\ and\ \citenamefont {Fern{\'a}ndez-Rossier}}]{ortiz18}%
  \BibitemOpen
  \bibfield  {author} {\bibinfo {author} {\bibfnamefont {R.}~\bibnamefont {Ortiz}}, \bibinfo {author} {\bibfnamefont {N.~A.}\ \bibnamefont {Garc{\'\i}a-Mart{\'\i}nez}}, \bibinfo {author} {\bibfnamefont {J.}~\bibnamefont {Lado}},\ and\ \bibinfo {author} {\bibfnamefont {J.}~\bibnamefont {Fern{\'a}ndez-Rossier}},\ }\href@noop {} {\bibfield  {journal} {\bibinfo  {journal} {Physical Review B}\ }\textbf {\bibinfo {volume} {97}},\ \bibinfo {pages} {195425} (\bibinfo {year} {2018})}\BibitemShut {NoStop}%
\bibitem [{\citenamefont {Fishman}\ \emph {et~al.}(2022)\citenamefont {Fishman}, \citenamefont {White},\ and\ \citenamefont {Stoudenmire}}]{ITensor}%
  \BibitemOpen
  \bibfield  {author} {\bibinfo {author} {\bibfnamefont {M.}~\bibnamefont {Fishman}}, \bibinfo {author} {\bibfnamefont {S.~R.}\ \bibnamefont {White}},\ and\ \bibinfo {author} {\bibfnamefont {E.~M.}\ \bibnamefont {Stoudenmire}},\ }\href {https://doi.org/10.21468/SciPostPhysCodeb.4} {\bibfield  {journal} {\bibinfo  {journal} {SciPost Phys. Codebases}\ ,\ \bibinfo {pages} {4}} (\bibinfo {year} {2022})}\BibitemShut {NoStop}%
\bibitem [{\citenamefont {Lado}(2023)}]{lado2023dmrgpy}%
  \BibitemOpen
  \bibfield  {author} {\bibinfo {author} {\bibfnamefont {J.}~\bibnamefont {Lado}},\ }\href {https://github.com/joselado/dmrgpy} {\bibinfo {title} {{DMRGpy library}}} (\bibinfo {year} {2023})\BibitemShut {NoStop}%
\bibitem [{\citenamefont {del Castillo}\ and\ \citenamefont {Fern{\'a}ndez-Rossier}(2024)}]{del24}%
  \BibitemOpen
  \bibfield  {author} {\bibinfo {author} {\bibfnamefont {Y.}~\bibnamefont {del Castillo}}\ and\ \bibinfo {author} {\bibfnamefont {J.}~\bibnamefont {Fern{\'a}ndez-Rossier}},\ }\href@noop {} {\bibfield  {journal} {\bibinfo  {journal} {Physical Review B}\ }\textbf {\bibinfo {volume} {110}},\ \bibinfo {pages} {045145} (\bibinfo {year} {2024})}\BibitemShut {NoStop}%
\bibitem [{sup()}]{supp}%
  \BibitemOpen
  \href@noop {} {}\bibinfo {note} {See Supplemental Material for a comparison between the complete Hamiltonian and the effective model, an estimation of the ground-state thermal occupation, and a discussion of possible experimental realizations of remote spin control. The Supplemental Material also contains Ref. [45-47].}\BibitemShut {Stop}%
\bibitem [{\citenamefont {Baumann}\ \emph {et~al.}(2015)\citenamefont {Baumann}, \citenamefont {Paul}, \citenamefont {Choi}, \citenamefont {Lutz}, \citenamefont {Ardavan},\ and\ \citenamefont {Heinrich}}]{baumann15}%
  \BibitemOpen
  \bibfield  {author} {\bibinfo {author} {\bibfnamefont {S.}~\bibnamefont {Baumann}}, \bibinfo {author} {\bibfnamefont {W.}~\bibnamefont {Paul}}, \bibinfo {author} {\bibfnamefont {T.}~\bibnamefont {Choi}}, \bibinfo {author} {\bibfnamefont {C.~P.}\ \bibnamefont {Lutz}}, \bibinfo {author} {\bibfnamefont {A.}~\bibnamefont {Ardavan}},\ and\ \bibinfo {author} {\bibfnamefont {A.~J.}\ \bibnamefont {Heinrich}},\ }\href@noop {} {\bibfield  {journal} {\bibinfo  {journal} {Science}\ }\textbf {\bibinfo {volume} {350}},\ \bibinfo {pages} {417} (\bibinfo {year} {2015})}\BibitemShut {NoStop}%
\bibitem [{\citenamefont {Maiellaro}\ \emph {et~al.}(2024)\citenamefont {Maiellaro}, \citenamefont {Aubin}, \citenamefont {Mesaros},\ and\ \citenamefont {Simon}}]{maiellaro24}%
  \BibitemOpen
  \bibfield  {author} {\bibinfo {author} {\bibfnamefont {A.}~\bibnamefont {Maiellaro}}, \bibinfo {author} {\bibfnamefont {H.}~\bibnamefont {Aubin}}, \bibinfo {author} {\bibfnamefont {A.}~\bibnamefont {Mesaros}},\ and\ \bibinfo {author} {\bibfnamefont {P.}~\bibnamefont {Simon}},\ }\href@noop {} {\bibfield  {journal} {\bibinfo  {journal} {Physical Review B}\ }\textbf {\bibinfo {volume} {110}},\ \bibinfo {pages} {L220410} (\bibinfo {year} {2024})}\BibitemShut {NoStop}%
\bibitem [{\citenamefont {Landau}(1932)}]{Landau1932}%
  \BibitemOpen
  \bibfield  {author} {\bibinfo {author} {\bibfnamefont {L.}~\bibnamefont {Landau}},\ }\href@noop {} {\bibfield  {journal} {\bibinfo  {journal} {Phys. Z. Sowjetunion}\ }\textbf {\bibinfo {volume} {2}},\ \bibinfo {pages} {118} (\bibinfo {year} {1932})}\BibitemShut {NoStop}%
\bibitem [{\citenamefont {Zener}(1932)}]{Zener1932}%
  \BibitemOpen
  \bibfield  {author} {\bibinfo {author} {\bibfnamefont {C.}~\bibnamefont {Zener}},\ }\href@noop {} {\bibfield  {journal} {\bibinfo  {journal} {Proceedings of the Royal Society of London. Series A, Containing Papers of a Mathematical and Physical Character}\ }\textbf {\bibinfo {volume} {137}},\ \bibinfo {pages} {696} (\bibinfo {year} {1932})}\BibitemShut {NoStop}%
\bibitem [{\citenamefont {Ivakhnenko}\ \emph {et~al.}(2023)\citenamefont {Ivakhnenko}, \citenamefont {Shevchenko},\ and\ \citenamefont {Nori}}]{ivakhnenko2023}%
  \BibitemOpen
  \bibfield  {author} {\bibinfo {author} {\bibfnamefont {V.}~\bibnamefont {Ivakhnenko}}, \bibinfo {author} {\bibfnamefont {S.~N.}\ \bibnamefont {Shevchenko}},\ and\ \bibinfo {author} {\bibfnamefont {F.}~\bibnamefont {Nori}},\ }\href@noop {} {\bibfield  {journal} {\bibinfo  {journal} {Physics Reports}\ }\textbf {\bibinfo {volume} {995}},\ \bibinfo {pages} {1} (\bibinfo {year} {2023})}\BibitemShut {NoStop}%
\bibitem [{\citenamefont {Rodr{\'\i}guez}\ \emph {et~al.}(2025)\citenamefont {Rodr{\'\i}guez}, \citenamefont {G{\'o}mez}, \citenamefont {Fern{\'a}ndez-Rossier},\ and\ \citenamefont {Ferr{\'o}n}}]{rodriguez25}%
  \BibitemOpen
  \bibfield  {author} {\bibinfo {author} {\bibfnamefont {S.~A.}\ \bibnamefont {Rodr{\'\i}guez}}, \bibinfo {author} {\bibfnamefont {S.~S.}\ \bibnamefont {G{\'o}mez}}, \bibinfo {author} {\bibfnamefont {J.}~\bibnamefont {Fern{\'a}ndez-Rossier}},\ and\ \bibinfo {author} {\bibfnamefont {A.}~\bibnamefont {Ferr{\'o}n}},\ }\href@noop {} {\bibfield  {journal} {\bibinfo  {journal} {Physical Review Letters}\ }\textbf {\bibinfo {volume} {134}},\ \bibinfo {pages} {056703} (\bibinfo {year} {2025})}\BibitemShut {NoStop}%
\end{thebibliography}%


\begin{thebibliography}{5}%
\makeatletter
\providecommand \@ifxundefined [1]{%
 \@ifx{#1\undefined}
}%
\providecommand \@ifnum [1]{%
 \ifnum #1\expandafter \@firstoftwo
 \else \expandafter \@secondoftwo
 \fi
}%
\providecommand \@ifx [1]{%
 \ifx #1\expandafter \@firstoftwo
 \else \expandafter \@secondoftwo
 \fi
}%
\providecommand \natexlab [1]{#1}%
\providecommand \enquote  [1]{``#1''}%
\providecommand \bibnamefont  [1]{#1}%
\providecommand \bibfnamefont [1]{#1}%
\providecommand \citenamefont [1]{#1}%
\providecommand \href@noop [0]{\@secondoftwo}%
\providecommand \href [0]{\begingroup \@sanitize@url \@href}%
\providecommand \@href[1]{\@@startlink{#1}\@@href}%
\providecommand \@@href[1]{\endgroup#1\@@endlink}%
\providecommand \@sanitize@url [0]{\catcode `\\12\catcode `\$12\catcode `\&12\catcode `\#12\catcode `\^12\catcode `\_12\catcode `\%12\relax}%
\providecommand \@@startlink[1]{}%
\providecommand \@@endlink[0]{}%
\providecommand \url  [0]{\begingroup\@sanitize@url \@url }%
\providecommand \@url [1]{\endgroup\@href {#1}{\urlprefix }}%
\providecommand \urlprefix  [0]{URL }%
\providecommand \Eprint [0]{\href }%
\providecommand \doibase [0]{https://doi.org/}%
\providecommand \selectlanguage [0]{\@gobble}%
\providecommand \bibinfo  [0]{\@secondoftwo}%
\providecommand \bibfield  [0]{\@secondoftwo}%
\providecommand \translation [1]{[#1]}%
\providecommand \BibitemOpen [0]{}%
\providecommand \bibitemStop [0]{}%
\providecommand \bibitemNoStop [0]{.\EOS\space}%
\providecommand \EOS [0]{\spacefactor3000\relax}%
\providecommand \BibitemShut  [1]{\csname bibitem#1\endcsname}%
\let\auto@bib@innerbib\@empty
%</preamble>
\bibitem [{\citenamefont {Mishra}\ \emph {et~al.}(2021)\citenamefont {Mishra}, \citenamefont {Catarina}, \citenamefont {Wu}, \citenamefont {Ortiz}, \citenamefont {Jacob}, \citenamefont {Eimre}, \citenamefont {Ma}, \citenamefont {Pignedoli}, \citenamefont {Feng}, \citenamefont {Ruffieux}, \citenamefont {Fernandez-Rossier},\ and\ \citenamefont {Fasel}}]{Mishra2021}%
  \BibitemOpen
  \bibfield  {author} {\bibinfo {author} {\bibfnamefont {S.}~\bibnamefont {Mishra}}, \bibinfo {author} {\bibfnamefont {G.}~\bibnamefont {Catarina}}, \bibinfo {author} {\bibfnamefont {F.}~\bibnamefont {Wu}}, \bibinfo {author} {\bibfnamefont {R.}~\bibnamefont {Ortiz}}, \bibinfo {author} {\bibfnamefont {D.}~\bibnamefont {Jacob}}, \bibinfo {author} {\bibfnamefont {K.}~\bibnamefont {Eimre}}, \bibinfo {author} {\bibfnamefont {J.}~\bibnamefont {Ma}}, \bibinfo {author} {\bibfnamefont {C.~A.}\ \bibnamefont {Pignedoli}}, \bibinfo {author} {\bibfnamefont {X.}~\bibnamefont {Feng}}, \bibinfo {author} {\bibfnamefont {P.}~\bibnamefont {Ruffieux}}, \bibinfo {author} {\bibfnamefont {J.}~\bibnamefont {Fernandez-Rossier}},\ and\ \bibinfo {author} {\bibfnamefont {R.}~\bibnamefont {Fasel}},\ }\href@noop {} {\bibfield  {journal} {\bibinfo  {journal} {Nature}\ }\textbf {\bibinfo {volume} {598}},\ \bibinfo {pages} {287} (\bibinfo {year} {2021})}\BibitemShut {NoStop}%
\bibitem [{\citenamefont {Zhao}\ \emph {et~al.}(2024)\citenamefont {Zhao}, \citenamefont {Catarina}, \citenamefont {Zhang}, \citenamefont {Henriques}, \citenamefont {Yang}, \citenamefont {Ma}, \citenamefont {Feng}, \citenamefont {Gr{\"o}ning}, \citenamefont {Ruffieux}, \citenamefont {Fern{\'a}ndez-Rossier} \emph {et~al.}}]{zhao24}%
  \BibitemOpen
  \bibfield  {author} {\bibinfo {author} {\bibfnamefont {C.}~\bibnamefont {Zhao}}, \bibinfo {author} {\bibfnamefont {G.}~\bibnamefont {Catarina}}, \bibinfo {author} {\bibfnamefont {J.-J.}\ \bibnamefont {Zhang}}, \bibinfo {author} {\bibfnamefont {J.~C.}\ \bibnamefont {Henriques}}, \bibinfo {author} {\bibfnamefont {L.}~\bibnamefont {Yang}}, \bibinfo {author} {\bibfnamefont {J.}~\bibnamefont {Ma}}, \bibinfo {author} {\bibfnamefont {X.}~\bibnamefont {Feng}}, \bibinfo {author} {\bibfnamefont {O.}~\bibnamefont {Gr{\"o}ning}}, \bibinfo {author} {\bibfnamefont {P.}~\bibnamefont {Ruffieux}}, \bibinfo {author} {\bibfnamefont {J.}~\bibnamefont {Fern{\'a}ndez-Rossier}}, \emph {et~al.},\ }\href@noop {} {\bibfield  {journal} {\bibinfo  {journal} {Nature Nanotechnology}\ ,\ \bibinfo {pages} {1}} (\bibinfo {year} {2024})}\BibitemShut {NoStop}%
\bibitem [{\citenamefont {Fishman}\ \emph {et~al.}(2022)\citenamefont {Fishman}, \citenamefont {White},\ and\ \citenamefont {Stoudenmire}}]{ITensor}%
  \BibitemOpen
  \bibfield  {author} {\bibinfo {author} {\bibfnamefont {M.}~\bibnamefont {Fishman}}, \bibinfo {author} {\bibfnamefont {S.~R.}\ \bibnamefont {White}},\ and\ \bibinfo {author} {\bibfnamefont {E.~M.}\ \bibnamefont {Stoudenmire}},\ }\href {https://doi.org/10.21468/SciPostPhysCodeb.4} {\bibfield  {journal} {\bibinfo  {journal} {SciPost Phys. Codebases}\ ,\ \bibinfo {pages} {4}} (\bibinfo {year} {2022})}\BibitemShut {NoStop}%
\bibitem [{\citenamefont {Lado}(2023)}]{lado2023dmrgpy}%
  \BibitemOpen
  \bibfield  {author} {\bibinfo {author} {\bibfnamefont {J.}~\bibnamefont {Lado}},\ }\href {https://github.com/joselado/dmrgpy} {\bibinfo {title} {{DMRGpy library}}} (\bibinfo {year} {2023})\BibitemShut {NoStop}%
\bibitem [{\citenamefont {del Castillo}\ and\ \citenamefont {Fern{\'a}ndez-Rossier}(2024)}]{del24}%
  \BibitemOpen
  \bibfield  {author} {\bibinfo {author} {\bibfnamefont {Y.}~\bibnamefont {del Castillo}}\ and\ \bibinfo {author} {\bibfnamefont {J.}~\bibnamefont {Fern{\'a}ndez-Rossier}},\ }\href@noop {} {\bibfield  {journal} {\bibinfo  {journal} {Physical Review B}\ }\textbf {\bibinfo {volume} {110}},\ \bibinfo {pages} {045145} (\bibinfo {year} {2024})}\BibitemShut {NoStop}%
\end{thebibliography}%

\end{document}

\title{Supplemental material of "Remote spin control  in Haldane spin chains"}

\author{Y. del Castillo$^{1,2}$, A. Ferr\'on $^3$, J. Fern\'{a}ndez-Rossier$^1$\footnote{On permanent leave from Departamento de F\'{i}sica Aplicada, Universidad de Alicante, 03690 San Vicente del Raspeig, Spain}$^,$\footnote{joaquin.fernandez-rossier@inl.int} }

\affiliation{$^1$International Iberian Nanotechnology Laboratory (INL), Av. Mestre Jos\'{e} Veiga, 4715-330 Braga, Portugal }
\affiliation{$^2$Centro de F\'{i}sica das Universidades do Minho e do Porto, Universidade do Minho, Campus de Gualtar, 4710-057 Braga, Portugal }
\affiliation{$^3$Instituto de Modelado e Innovación Tecnológica (CONICET-UNNE) and Facultad de Ciencias Exactas, Naturales y Agrimensura, Universidad Nacional del Nordeste, Avenida Libertad 5400, W3404AAS Corrientes, Argentina. }

\date{\today}

\maketitle

\section{Comparison between the effective model and the full Hamiltonian}

In this section, we assess the validity of the effective Hamiltonian description (Eq. (9)) against the exact numerical diagonalization of the full Hamiltonians Eq.(1) and Eq.(2), the $S=1$ Haldane spin chain and the alternating exchange Heisenberg model (AEHM) with $S=1/2$, respectively. For both models, we compute, as a function of the local field $b$ applied to the first spin, the energies of the low-energy manifold and the local spin expectation value $\langle \psi_{\pm} | \hat S^z_i |\psi_{\pm} \rangle$.

\begin{figure}[h]
    \centering
    \includegraphics[width=0.65 \linewidth]{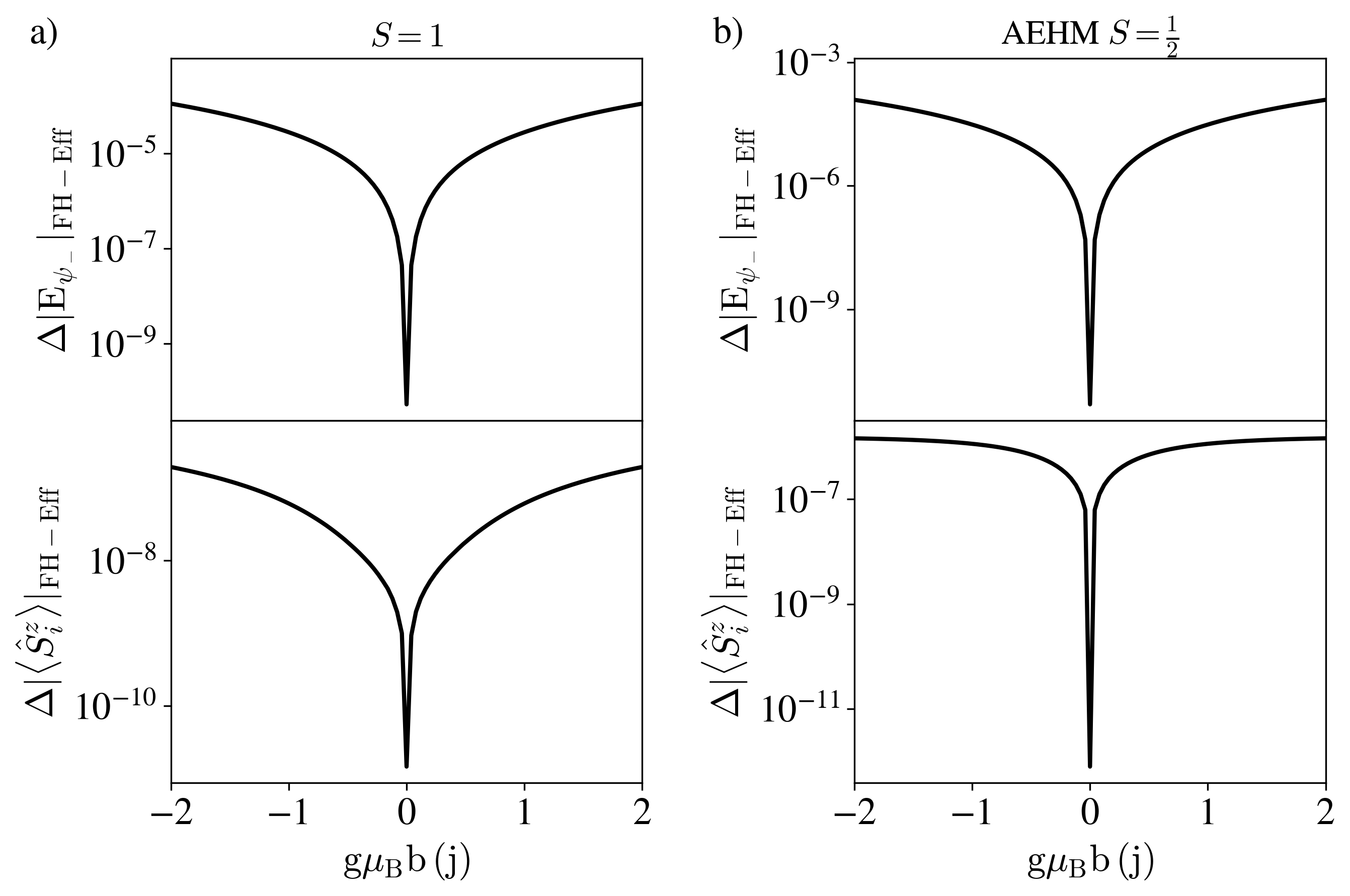}
    \caption{Comparison between the full Hamiltonian (FH) from Eq. (1) and (2) and the effective model (Eff) from Eq. (9). a) show the difference in the energies of the ground state, $\Delta E_{FH-Eff}$ and the difference in the local spin expectation value, $\Delta \langle \hat S^z_i \rangle_{FH-Eff}$, as a function of the local field $b$ applied to the first spin for the $S=1$ model. b) shows the same results for the AEHM model. Parameters for the $S=1$ model: $N=10$, $\beta=0.32$, Parameters for the AEHM $S=1/2$ model: N=12, $\delta$=-0.4.}
    \label{fig:SFIG1}
\end{figure}

Fig. \ref{fig:SFIG1} summarizes our results. Panel a) corresponds to the $S=1$ Haldane model, showing, respectively, the difference in energy for the ground state and the difference in $\langle S^z_i \rangle$ between the two models. Panel b) shows the same quantities for the AEHM. For the $S=1$ Haldane chain, the relative difference in energy is of the order of $10^{-4}$ for $g\mu_B b > j$, and the absolute difference in $\Delta \langle \hat S^z_i \rangle$ remains below $10^{-7}$ for the example. For the AEHM, the differences are slightly higher but of the same order of magnitude. We also note that there can be a mismatch due to numerical error in the ED calculation.

\section{Occupation of the ground state: energy scales}

For the effective Hamiltonian of Eq. (9) to remain valid, all the relevant energy scales, the singlet–triplet splitting $j$, the global magnetic field, the local magnetic field $b$, and the thermal energy $k_{\mathrm{B}}T$, must be much smaller than the Haldane gap $\Delta_{\mathrm{H}}$. This condition is naturally fulfilled for sufficiently long chains, since $j$ decreases exponentially with system size and  $\Delta_{\mathrm{H}}$ can be in the order of tens of meV in the case of nanographene-based chains \cite{Mishra2021,zhao24}.

In order to apply the dynamical sweeps discussed in the main text while keeping the system in its ground state, we must control the sweeping velocity, but additionally, we need to ensure that the ground state is also the most thermally occupied state within the four-level manifold. This requires $
    k_{\mathrm{B}} T \ll j $,
so that the thermal population of the excited states is negligible.  
Given a spectrum $\{E_n\}$, the ground state occupation probability at finite temperature is given by $P_{\mathrm{GS}} = \frac{e^{-E_{\psi_-}/k_{\mathrm{B}}T}}{Z}$, 
where $E_{\psi_-}$ is the ground state energy and $Z$ is the partition function.

\begin{figure}[h]
    \centering
    \includegraphics[width= 0.6\linewidth]{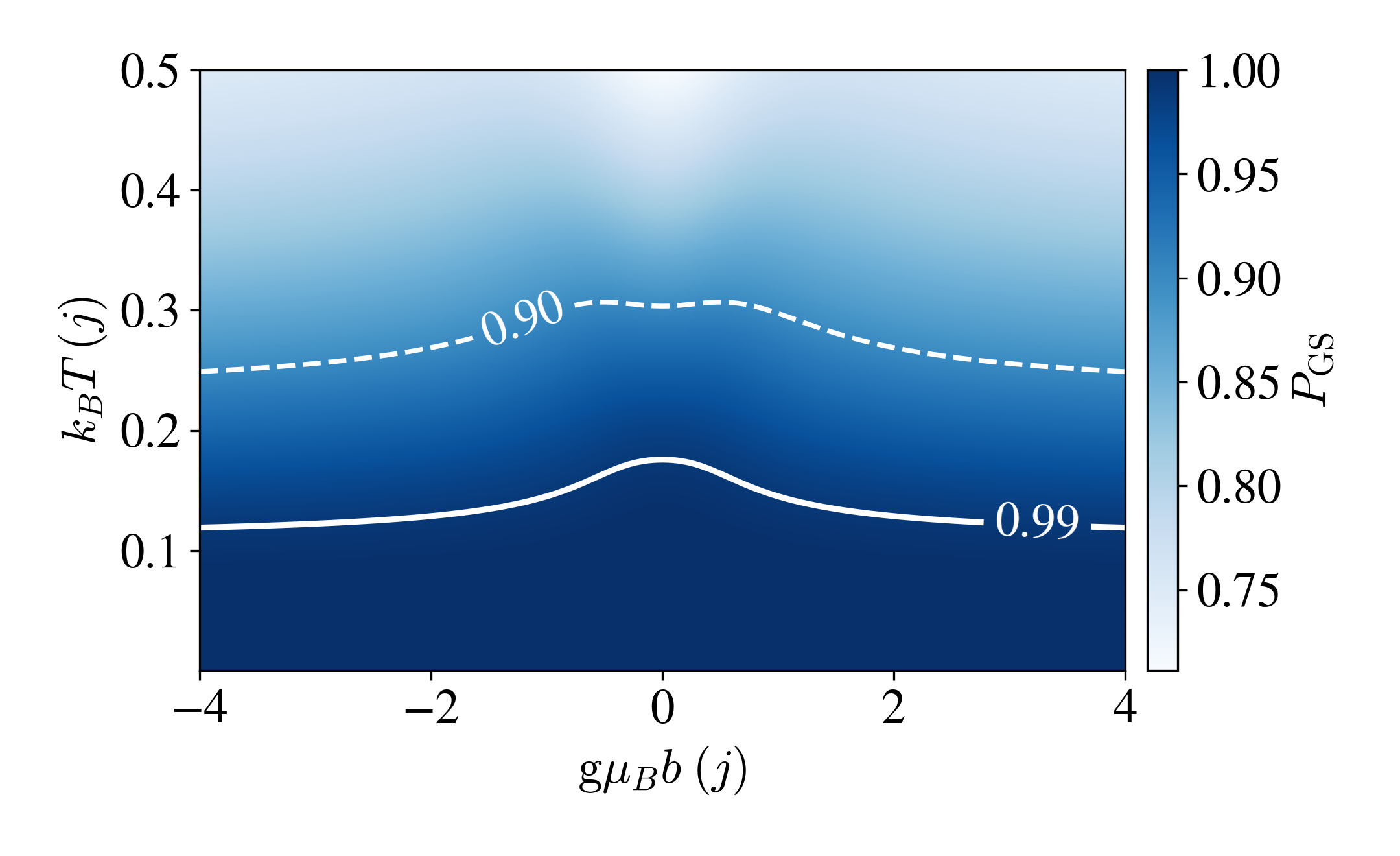}
    \caption{ Ground state thermal occupation probability $P_{\mathrm{GS}}$ as a function of local field $b$ and temperature $T$. Contour lines mark 90\% and 99\% occupation. Parameters: $N=12$, $\beta =0.32$.}
    \label{fig:SFIG2}
\end{figure}

Fig. \ref{fig:SFIG2} shows $P_{\mathrm{GS}}$ as a function of temperature and local field $b$, both expressed in units of $j$. The contour lines indicate 90\% and 99\% ground state occupation. As expected, $P_{\mathrm{GS}}$ decreases with increasing $T$, but for a fixed $T$, the occupation saturates for large values of $b$. This saturation comes from the constant energy separation between the state $\psi_{-}$ and one of the triplet states that is also sensitive to the local field. In our simulations, the convergence occurs for $k_{\mathrm{B}}T \simeq 0.11\, j$.

An external magnetic field applied uniformly to the chain does not affect the $\psi_{-}$ and $\psi_{+}$ states, since both have total $S^z = 0$. However, the Zeeman term is present in the $S^z = \pm 1$ states and would shift their energies linearly, affecting the thermal occupation of the ground state. Hence, the external field should be of the same order of magnitude as the thermal energy.

The local perturbation can be applied not only on one edge of the chain but on any spin. Possible solutions to overcome the effect of external magnetic fields can be used to boost the occupation of the ground state. For instance, two local perturbations with opposite signs at each edge of the chain can also be used to polarize the chain, dividing the total local field needed for the sweep between two sources of perturbation. In that arrangement, the local perturbations with opposite signs cancel each other’s effect on the triplet states with $S^z = \pm 1$, and hence the avoided crossing level is realizable while the $T_{\pm}$ states remain at the same energy. Their distance from the ground state increases, promoting its thermal occupation. 

Another solution is to use a controlled external field in real time, where the local perturbation can be matched at each step of the sweep with a similar external field of similar magnitude (half of the local perturbation in the case of a single local perturbation on one edge). This would leave the evolution of the energy in the $T_{\pm}$ states constant, again promoting the occupation of the ground state.

\section{Realizations of the controlled sweep}

In order to perform the adiabatic control of the edge spin, from one polarized state to the other, the sign of the local perturbation needs to change. Here, we propose different scenarios where this could be realized. 

The simplest scenario involves the use of an additional magnetic atom or molecule close to the chain with an almost classical behavior, for example, a Ho atom with a long lifetime for each state. This almost classical magnet can be positioned next to the chain at a distance where the stray field induced at the edge of the chain polarizes the chain and serves as the initial position for the sweep. Then, a magnetic STM tip, generating a stray field with opposite sign to the magnetic atom (or the same sign if the STM tip is on the other edge), can be brought close to the chain edge in a controlled manner, eventually overcoming the other stray field and changing the sign of the perturbation.

Another possible approach is to combine a local tip-induced field with a uniform background field from a ferromagnetic substrate, separated from the spin chain by a decoupling layer. On ferromagnetic surfaces, domain walls can be manipulated to change the local magnetization under the chain. The advantage of this method is that the ferromagnetic surface can generate very large local fields.

\section{Density Matrix Renormalization Group (DMRG) computation details}

\begin{figure}[h]
    \centering
    \includegraphics[width=0.5\linewidth]{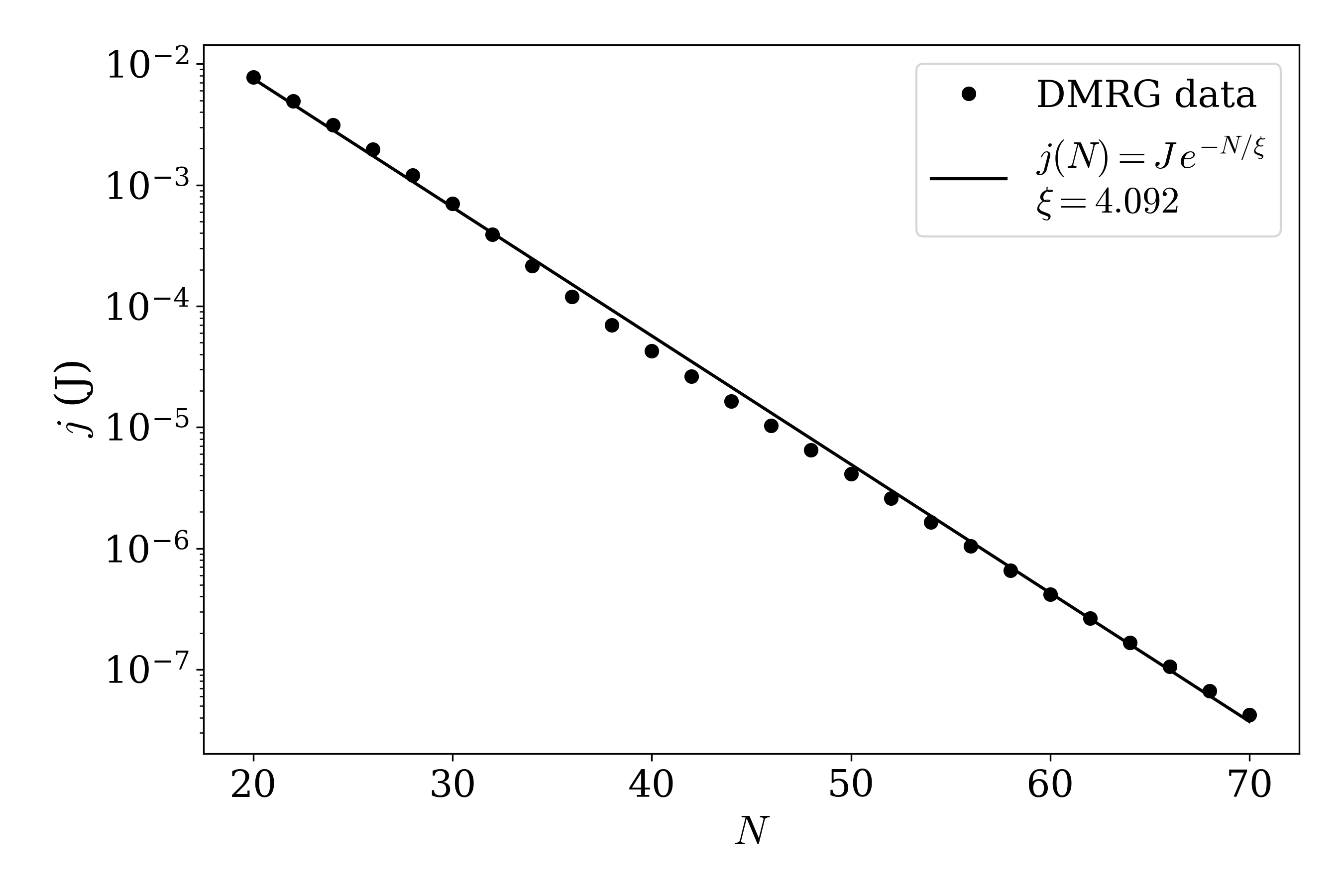}
    \caption{Singlet-triplet gap $j$ for the $S=1$ Haldane model, computed for different chain lengths, $N$.}
    \label{fig:SFIG3}
\end{figure}

We computed the singlet–triplet gap of the $S=1$ Haldane Hamiltonian and the $S=1/2$ AEHM with finite-size two-site DMRG, implemented with \texttt{ITensors.jl}\cite{ITensor}. For the range discussed in the examples ($N=20$ to $N=70$), the calculations used up to $n_{\mathrm{sweeps}}=30$ and a maximum bond dimension $m_{\max}=100$ with open boundary conditions and $S^z$ conservation. In Fig. $\ref{fig:SFIG3}$, we show our computed $j$ for different chain lengths, $N$. We fit the exponential decay to $ j \simeq J e^{-N/\xi}$ and we find a $\xi = 4.09 $

To obtain the non-vanishing matrix element ${\mathcal{S}_i}$, required to compute the mixing term in the effective model $\varepsilon_0$, one needs to calculate the overlap between the singlet and the $T_0$ states. This task can become challenging within DMRG for large systems. However, the relation $|\mathcal{S}_i| = |\mathcal{T}_{i}^{\pm}|$ allows us to obtain this value in a less computationally demanding way.  

In practice, during the DMRG calculation, we artificially promote one of the $S^z=\pm1$ triplet states by introducing a Zeeman term. With a well-defined ground state, the DMRG simulation converges faster and more accurately. Once the local magnetic moments of each spin are obtained, it is sufficient to flip the sign of half of the chain to recover exactly the value of $\mathcal{S}_i$. These elements were obtained with \texttt{dmrgpy} \cite{lado2023dmrgpy}. Experimentally, it should also be possible to get the unperturbed magnetic moments of a Haldane chain using electron spin resonance based on scanning tunneling microscopy magnetometry\cite{del24}.

\bibliographystyle{apsrev4-2}
\bibliography{biblio}{}